\documentclass[aps,prd,nofootinbib,reprint,superscriptaddress,longbibliography]{revtex4-1}
\usepackage{etoolbox}
\usepackage{dcolumn}
\usepackage{bm}
\usepackage{graphics}
\usepackage{dirtytalk}
\usepackage{graphicx} 
\usepackage{blindtext}
\usepackage{amsmath}
\usepackage{amsfonts}
\usepackage{amssymb}
\usepackage[%
colorlinks=true,
urlcolor=blue,
linkcolor=red,
citecolor=blue
]{hyperref}

\usepackage{wrapfig}

\usepackage{natbib}
\begin{document}
\title{\bf Investigating the role of mutual information in the Page curve for a functional renormalization group improved Schwarzschild black hole}
\vskip 1cm
	\vskip 1cm	
\author{Ashis Saha}
\email{ashis.saha@bose.res.in}
\author{Anirban Roy Chowdhury}
\email{iamanirban@bose.res.in}
\author{Sunandan Gangopadhyay}
\email{sunandan.gangopadhyay@bose.res.in}
\affiliation{Department of Astrophysics and High Energy physics\linebreak
	S.N.~Bose National Centre for Basic Sciences, JD Block, Sector-III, Salt Lake, Kolkata 700106, India}
	\begin{abstract}
	\noindent The present work delves into probing the importance of mutual information of relevant subsystems in obtaining the correct time-evolution of fine-grained entropy of Hawking radiation, as suggested by Page. This was done by considering a functional renormalization group improved or simply, quantum corrected Schwarzschild black solution which captures the flavour of an effective theory of quantum gravity. The mentioned black hole solution emerges from an asymptotically safe average effective action which describes a trajectory in momentum-space and satisfies a renormalizarion group equation. Furthermore, in the before Page time scenario, the behaviour of the mutual information between appropiate subsystems over time leads to the Hartman-Maldacena time. The observations made for the quantum corrected Schwarzschild black hole have been compared to the same made for the standard Schwarzschild black hole in order to draw some novel conclusions. Based upon our observations, we also propose a formula for computing fine-grained entropy of Hawking radiation for an eternal black hole, in presence of the island.
	\end{abstract}
\maketitle
\noindent 
Black hole information paradox has emerged as one of the most fascinating topic of research in modern day physics. Hawking, in his seminal work \cite{Hawking:1974rv}, have shown that particle-antiparticle pair production in the near-horizon region of a black hole can be understood as the result of the vacuum fluctuations of the quantum fields. Furthermore, the presence of strong tidal force in this regime (near-horizon) can separate these pairs, leading us to one of the most remarkable and mysterious phenomenon in theoretical physics, known as the Hawking radiation of black holes. Although the corresponding calculations are originally semi-classical \cite{Hawking:1974rv,Hawking:1975vcx} but years of investigations in this direction has led us to the realization that this phenomena is purely quantum mechanical and one needs to incorporate the proper theory of quantum gravity to reveal the complete picture of Hawking radiation. However, we still do not have a completely reliable theory of quantum gravity. Nevertheless, there are certain beautiful aspects associated to this phenomenon which we have been able to disclose based upon our understanding of gravitation so far. Among these aspects, black hole thermodynamics actually pop up quite naturally as the black hole radiation appears as a thermal spectrum (for an observer at asymptotic infinity), attended by a finite temperature, commonly known as the Hawking temperature \cite{Hawking:1975vcx}. The most interesting finding of black hole thermodynamics is the entropy of a black hole which we generally denote as the Bekenstein-Hawking entropy of a black hole ($S_{BH}$) \cite{Bekenstein:1973ur} which tells us that the Boltzmann entropy of a black hole is non-extensive in its nature as it is proportional to the area of the event-horizon of the black hole. Further, as a familiar expectation, this entropy also corresponds to a set of thermodynamic laws \cite{Bekenstein:1972tm,PhysRevD.9.3292,Bardeen:1973gs}. A nice review in this regard can be found in \cite{Wald:1999vt}.\\
However, as the origin of the Hawking radiation relies upon the fundamental concepts of quantum physics, it will be a matter of basic instinct to probe the relevant measurements of quantum statistical mechanics also. Historically, this quest has Hawking to the computation of von-Neumann entropy \cite{Chuang:2000} of the black hole radiation which is a fine-grained notion of entropy unlike the usual coarse-grained notion of entropy (Bekenstein-Hawking entropy). In \cite{PhysRevD.14.2460}, he showed that the von Neumann entropy of Hawking radiation is a monotonically increasing function with respect to time of an asymptotic observer but from the landscape of quantum mechanical postulates, this statement creates a paradoxical scenario. This can be understood in the following way. It is known that the formation of black holes are (due to the collapse of a massive shell) corresponds to a pure state which is represented by the fact that at the beginning of the evaporation process, the von Neumann entropy of radiation is zero. Moreover, at the end of the evaporation process, the von Neumann entropy of radiation once again should be zero. This is quite a expected result from the perspective of unitary evolution of the quantum state over time. This contradictory situation can be denoted as the entropic version of the information paradox and the best way to understand the scenario is to take help of the Page curve. The Page curve describes the time evolution of the von Neumann entropy of Hawking radiation \cite{PhysRevLett.71.3743}. One can also naively say that this paradoxical situation can be solved if one can obtain the Page curve of radiation which abide by the rules of quantum mechanics. Before we indulge in to the subtleties of the Page curve it should be mentioned that, the von Neumann entropy of the Hawking radiation is identified with that of the matter fields residing on the outside region of the black hole, namely, region $R$. Now, according to the Page curve, starting from zero, the von-Neumann entropy of the radiation $(S_{vN}(R))$ increases over time. On the other hand, due to the evaporation the thermal entropy of black hole decreases with time and at a particular time (the Page time $t_P$) it becomes equal to the thermal entropy of the black hole (that is the Bekenstein-Hawking entropy $S_{BH}$). This can also be depicted as fine-grained entropy achieving the coarse-grained or maximal value. It can be noted that upto Page time there is no paradoxical situation because in this time domain $(t\le t_{P})$, the fine-grained entropy of radiation is bounded by the coarse-grained (thermal) entropy of the black hole, that is $S_{vN}(R)\leq S_{BH}$. However, after the Page time, for $t> t_{P}$ the von Neumann entropy of Hawking radiation is greater than the thermal entropy of the black hole. Thus the paradox appears just after the Page time and therefore the challenge is to obtain the correct behaviour of fine-grained entropy after the Page time. As we have mentioned earlier, the unitary principle suggests that $S_{vN}(R)$ should decrease after the Page time and at the end of the evaporation it should vanish \cite{PhysRevLett.71.3743,Page_2013}. Some interesting observations in this directions can be found in \cite{Almheiri:2012rt,Almheiri:2013hfa,Lloyd:2013bza,Papadodimas:2013wnh}.\\
Recent advancements of the AdS/CFT correspondence has provided us a fresh insight to the above discussed scenario. It has been suggested that while computing the von Neumann entropy of radiation, we have to take into account the contribution of certain region formed inside black hole horizon along with the traditional contributions coming from the matter fields located outside the black hole, where the radiation is collected. This region which is formed in the interior of the black hole is known as the \textit{islands} \cite{Penington:2019kki,Penington:2019npb,Almheiri:2019hni,Almheiri:2019psf,Almheiri:2019qdq,Almheiri:2019yqk,Almheiri:2020cfm}, and the end points of the island surface is called the \textit{quantum extremal surfaces} (QES) \cite{Engelhardt:2014gca,Engelhardt:2019hmr,Akers:2019lzs,Wall:2012uf}. Keeping this additional contribution in mind, the fine-grained entropy of Hawking radiation in the presence of island is given by 
\begin{equation}\label{eq1}
S (R) =	\textrm{min}~\mathop{\textrm{ext}}_{\mathcal{\mathrm{I}}}\bigg\{\frac{\textrm{Area}(\partial I)}{4G_N}+S_{vN}(I\cup R)\bigg\}~.
\end{equation}
In a simple sense, the first term in the above formula represents the geometrical contributions  and the second term summarizes the contribution coming from the von Neumann entropy of matter fields localized on the region $R\cup I$. The main essence of the above formula lies in the fact that it provides us a non-trivial generalization of the fine-grained entropy $S_{vN}(R)$. It can be observed that in order to obtain the correct value of $S_{vN}(R)$, one has to perform extremization over all possible island configurations $I$ and has to the consider the minimum of all extremal values, if there is more than one extrema.\\
The island formula has emerged from the application of replica trick to compute the von-Neumann entropy of matter fields, in presence of a dynamical gravitational background. In this method, the partition function appears as a gravitational path integral. Apart from the previously known Hawking saddle-point, it has been noted that there exists an additional non-trivial saddle-point known as the replica wormhole saddle-point\footnote{A detailed review regarding wormholes and their utility in gauge/gravity duality can be found in \cite{Kundu:2021nwp}.} (with proper boundary conditions) which acts as the dominating saddle-point depending under certain condition \cite{Almheiri:2019qdq,Penington:2019kki,Almheiri:2019qdq,Goto:2020wnk,Colin-Ellerin:2020mva}. This new formula of computing the fine-grained of Hawking radiation produces the Page curve which one expects from the principle of unitarity and due to this remarkable observation, the island formulation has emerged as an important prescription to be studied in recent times \cite{Chen:2019uhq,Hashimoto:2020cas,Hartman:2020swn,Anegawa:2020ezn,Dong:2020uxp,Balasubramanian:2020xqf,Alishahiha:2020qza,Azarnia:2021uch,Arefeva:2021kfx,He:2021mst,Omidi:2021opl,Yu:2021rfg,Yadav:2022fmo,Du:2022vvg}. However, there are certain issues regarding this approach which needs to be properly addressed \cite{Geng:2020qvw,Geng:2021hlu}.\\
This formalism was first applied to the black hole solution of the JT gravity filled with 2d conformal matter \cite{Almheiri:2019hni,Hollowood:2020cou,Goto:2020wnk}. In early time scenario, that is, $t\le t_{P}$, the Hawking saddle-point of the gravitational path integral dominates and the fine-grained entropy of Hawking radiation increases with respect to time. At the Page time $t=t_P$, it becomes equal to the value of coarse-grained entropy of radiation. However, just after the Page time, that is for $t>t_{P}$, the replica wormhole saddle-point starts to dominate and the contribution from the interior region ($I$) becomes significant which eventually lead us to the expected Page curve. There
are many interesting developments in recent years in this directions, such as 2d asymptotically flat or AdS black
holes \cite{Anegawa:2020ezn,Hartman:2020swn, Wang:2021mqq,Li:2021lfo,Pedraza:2021cvx}, higher dimensional black holes \cite{Almheiri:2019psy,He:2021mst,Alishahiha:2020qza,Hashimoto:2020cas,Matsuo:2020ypv,Krishnan:2020fer,Krishnan:2020oun,Omidi:2021opl,Geng:2021iyq,Geng:2021mic,Goswami:2022ylc,Du:2022vvg,Chu:2021gdb},
charged black holes \cite{Ling:2020laa,Wang:2021woy,Kim:2021gzd,Ahn:2021chg,Yu:2021cgi} etc.  Some other interesting studies in this direction can also be found in \cite{Balasubramanian:2020xqf,Chen:2020tes,Azarnia:2021uch,Geng:2021wcq,Bhattacharya:2021jrn,Yu:2024fks,Gomez:2024fij,Jain:2023xta,Guo:2023fly,2699316,Anand:2023ozw,Chang:2023gkt,Li:2023fly,Guo:2023gfa,Miao:2023unv,Yadav:2022jib,HosseiniMansoori:2022hok,Hu:2022zgy,Yu:2022xlh,Goswami:2022ylc,Djordjevic:2022qdk,Anand:2022mla,Azarnia:2022kmp,Seo:2022ezk,Yu:2024fks,Wu:2024jgy,Yu:2023whl,Luongo:2023jyz,Uhlemann_2021,Karch_2022}.\\
\noindent In \cite{Saha:2021ohr}, it was first shown that the effectiveness of the island contributions to fine-grained entropy of radiation can be firmly quantified in terms of the mutual information of the matter fields. To be precise, just after the Page time, the significant contributions coming from the island leads to the exact vanishing of the mutual information between the matter fields which are localized on the relevant regions. Furthermore, this saturation of mutual information surprisingly gifts us the \textit{Scrambling time} and by utilizing it one obtains the desired Page curve for an eternal black hole. This observation was further pursued in \cite{RoyChowdhury:2022awr,RoyChowdhury:2023eol} and also extended to the before Page time scenario where it was shown that from the behaviour of mutual information (between the matter fields localized on concerned regions) with respect to observer's time, one can find out the Hartman-Maldacena time in this landscape.\\
In this work, we have considered a quantum corrected Schwarzschild black hole solution to study the above discussed studies regarding mutual information in the framework of island. This solution emerges from the \textit{asymptotically safe} idea for constructing a effective theory of quantum gravity \cite{Reuter:1996cp,doi:10.1142/10369,reuter2019quantum} which was first proposed by Wienberg \cite{Weinberg:1980gg}. The motivation to consider the mentioned geometry is pretty straight forward, that is, we want to explore the proposals given in \cite{Saha:2021ohr,RoyChowdhury:2022awr,RoyChowdhury:2023eol} for such a gravitational solution which has the essence of quantum gravity.\\
The paper is organized in the following way. In section \eqref{Sec1}, we briefly discuss the necessary details regarding the black hole geometry under consideration. Our main studies can be found in section \eqref{Sec2} (after Page time scenario) and in section \eqref{Sec3} (before Page time scenario). Finally, we summarize our findings in section \eqref{Sec4}.

\section{Asymptotically safe theory of quantum gravity and the quantum corrected Schwarzschild black hole}\label{Sec1}
\begin{figure}[h!]
	\centering
	\includegraphics[scale=1]{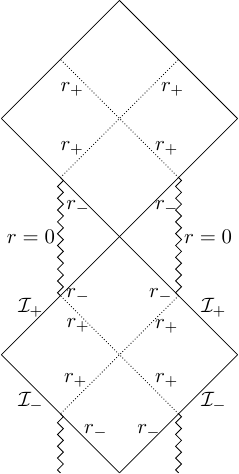}
	\caption{Penrose diagram of a two-sided eternal \textit{quantum corrected} eternal Schwarzschild black hole geometry.}
	\label{Fig1}
\end{figure}

The quest for a proper theory of quantum gravity has always been a matter of utmost interest due to the obvious reasons. Our understating of gravitation so far suggests that the standard perturbative approach (which is a traditional approach to construct a renormalizable quantum gauge theory) fails miserably for gravity as it leads to infinite number of ultraviolet (UV) divergences which require infinite number of distinct counter terms from the perspective of renormalization. This in turn means that only way out of this is a non-perturbative approach such as loop quantum gravity or considering gravity as a part of a more bigger picture \`a la supergravity or perhaps string theory is the final answer. However, there exists a different approach known as the functional renormalization group (RG) approach of quantum gravity in which one has to construct a scale-dependent RG equation which includes all possible diffeomorphism invariant (local) functions of the metric \cite{Reuter:1996cp,Reuter:2001ag}. Solution of this RG equation (described in terms of the momentum scale parameter $k$) leads to the low energy effective action \cite{Wetterich:1992yh,Reuter:1993kw,Lauscher:2002sq,PhysRevLett.103.101303,Mandal:2019xlg,Mandal:2020umo}. This results in a scale-dependent running nature of the dimensionful Newton's gravitational constant and by using this running coupling constant one \textit{improves} the Einstein equation. This effective theory constructed from this formulation has a non-trivial fixed point of the RG flow associated to its bare action. Furthermore, as there exists a UV fixed points one generally denotes the resultant theory as \textit{asymptotically safe} \cite{Weinberg:1980gg,Niedermaier:2006ns,Niedermaier:2003fz,Nink:2012vd,Litim:2014uca}.\\
In this work, we use a formulation of asymptotically safe gravity  based on an (Euclidean) ``effective average action" $\Gamma_k[g_{\mu\nu}]$ which correctly describes all gravitational
phenomena, including the effect of all loops, at a momentum scale $k$ \cite{Reuter:1996cp}. The effective average action comes out of modified theory in which the total action consists of the bare action, a regulator term to suppress all infrared (IR) modes with momenta $p^2<k^2$ below a chosen cut-off $k$. This in turn means all the modes with $p^2<k^2$ are excluded and the modes with $p^2>k^2$ are integrated out. As a result, the effective average action $\Gamma_k$ interpolates between the classical action $S_{cl}\equiv \Gamma_{k\rightarrow\infty}$ and the standard quantum effective action $S_{q}\equiv \Gamma_{k\rightarrow0}$ and it describes a trajectory in $k$-space which satisfies a RG flow equation, as mentioned earlier. In order to solve the RG equation, we consider the following form of the effective average action \cite{Reuter:1996cp}
\begin{eqnarray}
\Gamma_{k}[g,\bar{g}]=-\frac{1}{16\pi G(k)}\int d^dx \sqrt{g}~ \mathcal{R}(g)+S_{gauge}[g,\bar{g}]\nonumber\\
\end{eqnarray}
where $\mathcal{R}$ is the Ricci scalar, $\bar{g}_{\mu\nu}$ is the background metric tensor and $S_{gauge}[g,\bar{g}]$ is a classical gauge fixing term. The above form is a truncated form of the effective action which can be obtained from the method of \textit{truncation in theory space}. The RG equation corresponding to the above action leads to the following flow of Newton's constant
\begin{eqnarray}\label{eqRG}
	G(k)= \frac{G_0}{1+\omega G_0 k^2}
\end{eqnarray} 
where $G_0$ denotes the Newton's constant at $k=0$ and $\omega$ is a constant which has a quantum mechanical origin as it originates from the functional RG equation.\\
On the other hand, the Schwarzschild black hole metric is the classical solution of the Einstein field equation in the absence of any matter field. The spacetime metric can be written as 
\begin{equation}
	ds^2=-\left(1-\frac{2G_0M}{r}\right)dt^{2}+\frac{dr^{2}}{\left(1-\frac{2G_0M}{r}\right)}+r^{2}d\Omega_2^2
\end{equation}
where $G_0$ is the Newton's gravitational constant and $M$ represents the mass of the black hole. The Schwarzschild solution is very useful to describe many physical phenomena. This solution of Einstein equation is purely classical. The horizon radius of the Schwarzschild black hole is given by $r^{(S)}=2G_0M$.
The Hawking temperature $(T_{H}^{(S)})$ and the entropy $(S_{BH}^{(S)})$ of the Schwarzschild black hole are given by
\begin{eqnarray}\label{Eq3}
	T_{H}^{(S)}&=&\frac{1}{4\pi r^{(S)}}=\frac{1}{8\pi G_0M}\\
	S_{BH}^{(S)}&=&\frac{4\pi (r^{(S)})^2}{4G_0}
\end{eqnarray}
In \cite{Bonanno:2000ep}, a RG improved or quantum corrected Schwarzschild black hole (QCSBH) spacetime was proposed by identifying $G(k)\equiv G(r)$ where the form of $G(k)$ is given in eq.\eqref{eqRG}. In order to do that, one needs to compute the position dependent form of Newton's constant $G(r)$ from $G(k)$. This is to be done by identifying the infrared cut-off scale of $k$. For the choice $k=\frac{\xi}{d(r)}$ of the cut-off scale \cite{Bonanno:2000ep}, one gets
\begin{eqnarray}\label{eqG}
	G(r)=\frac{G_0}{1+\omega G_0 \left(\frac{\xi}{d(r)}\right)^2}
	\equiv \frac{G_0 d^2(r)}{d^2(r)+\tilde{\omega} G_0}
\end{eqnarray}
where $\tilde{\omega}=\omega \xi^2$ and $d(r)$ can be interpreted as the proper distance from the point $P$ (in Schwarzschild coordinate ($t,r,\theta,\phi$)) to the centre of the black hole along some curve. In \cite{Bonanno:2000ep}, the expression for $d(r)$ was obtained to be 
\begin{eqnarray}
	d(r)=\sqrt{\frac{r^3}{r+\gamma G_0 M^2}}~.
\end{eqnarray}
Now, we make use of the above form of $d(r)$ in eq.\eqref{eqG} to obtain the following result \footnote{Here, we are neglecting the term $\sim\mathcal{O}(G_0^2)$ as the contribution coming from this term is very small.}\cite{Mandal:2022quv} 
\begin{eqnarray}
G(r)=\frac{G_0}{1+\frac{\tilde{\omega} G_0}{r^2}+\frac{\tilde{\omega}\gamma G_0^2 M}{r^3}} \approxeq \frac{G_0}{1+\frac{\tilde{\omega} G_0}{r^2}}\nonumber\\~.
\end{eqnarray}
This in turn means we can get the spacetime metric of RG improved or quantum corrected Schwarzschild black hole solution by using the above obtained form of $G(r)$. This reads
\begin{eqnarray}\label{Eq5}
ds^2=-f(r)dt^{2}+\frac{dr^2}{f(r)}+r^{2} d\Omega_2^2
\end{eqnarray}
where the lapse function $f(r)$ is given by 
\begin{equation}
f(r)=1-\frac{2G(r)M}{r}~,~G(r)=\frac{G_0}{1+\frac{\tilde{\omega }G_0}{r^2}}
\end{equation}
where $\tilde{\omega}$ is a constant which represents the quantum correction to the classical black hole solution. The horizon radius can be obtained by using the condition $f(r)=0$, which leads to the following results 
\begin{equation}
r_{\pm}=G_0M\pm\sqrt{G_0^2M^2-\tilde{\omega}G_0},
\end{equation}
where $r_{+}$ and $r_{-}$ represent the outer and inner horizons respectively. It is also important to note that the number of horizons depends sensitively on the mass parameter. In fact, it can be shown that there exists a critical mass \( M_{\text{cr}} \) that governs the horizon structure, for masses greater than \( M_{\text{cr}} \), the black hole exhibits two distinct horizons, while for masses below this threshold, no horizon exists, and if $M=M_{cr}$ there is only one horizon. In this work, we have considered the scenario where \( M > M_{\text{cr}} \), which implies the existence of two distinct horizons \cite{Koch:2014cqa}. Therefore,
we can rewrite the lapse function in terms of $r_{+}$ and $r_{-}$ in the following way
\begin{eqnarray}\label{lap}
f(r)=\frac{(r-r_{+})(r-r_{-})}{r^2+\tilde{\omega}G_{0}}~.
\end{eqnarray}
Now if we consider the quantum correction parameter $\tilde{\omega}$ to be small. Under this considerations, the outer horizon radius and the inner horizon radius simplifies as
\begin{eqnarray}\label{EQ8}
r_{+}&\approx&2G_0M-\frac{\tilde{\omega }}{2M}=r^{(S)}-\frac{\tilde{\omega }}{2M}\\
r_{-}&\approx&\frac{\tilde{\omega }}{2M}~.
\end{eqnarray}
In this work we are only interested in the black hole sector. 
For the subsequent analysis, we will use the above form of horizon radius. The surface gravity and then Hawking temperature of QCSBH \footnote{The inverse of $T_{H}^{(q)}$, $\beta^{(q)}=8\pi G_{0}M\left(1+\frac{\tilde{\omega}G_{0}}{(r^{(s)})^2}\right)=\beta^{(S)}\left(1+\frac{\tilde{\omega}G_{0}}{(r^{(s)})^2}\right)$, where $\beta^{(S)}=8\pi GM,$ is the inverse temperature of the classical Schwarzschild black hole.}is given by
\begin{eqnarray}
\kappa^{(q)}&=&\frac{1}{4G_0M}\left(1-\frac{\tilde{\omega}G_{0}}{(r^{(s)})^2}\right)\\
T_{H}^{(q)}&=&T_{H}^{(S)}\left(1-\frac{\tilde{\omega}G_{0}}{(r^{(s)})^2}\right)
\end{eqnarray}
where $T_{H}^{(S)}$ is the temperature of the classical Schwarzschild black hole  solution (given in eq.\eqref{Eq3}). The thermodynamic entropy of QCSBH is obtained to be
\begin{eqnarray}\label{EQ13}
S_{BH}^{(q)}=\frac{4\pi r_{+}^2}{4G_0}~.
\end{eqnarray}
We now write down the spacetime metric given in \eqref{Eq5} in terms of the Kruskal coordinates. This reads
\begin{eqnarray}\label{Eq12}
ds^2=-F^2(r)dUdV+r^2d\Omega^2~,~ F^2(r)=\frac{f(r)}{\kappa^{(q)^2}}e^{-2\kappa^{(q)}r^{*}(r)}\nonumber\\
\end{eqnarray}
where $r^{*}$ is the tortoise coordinate and $U,V$ are Kruskal coordinates. The transformation rule for the tortoise coordinate reads
\begin{eqnarray}\label{Eq13}
r^{*}(r)&=&\int \frac{dr}{f(r)}\nonumber\\
&=&r+\left(\frac{r_{+}^2+\tilde{\omega}G_{0}}{r_{+}-r_{-}}\right)\log(r-r_{+})\nonumber\\
&&-\left(\frac{r_{-}^2+\tilde{\omega}G_{0}}{r_{+}-r_{-}}\right)\log(r-r_{-})~.
\end{eqnarray}
The Kruskal coordinate transformation for the right wedge reads
\begin{eqnarray}
U&=&-e^{-\kappa^{(q)}(t-r^*(r))}\nonumber\\
V&=&e^{\kappa^{(q)}(t+r^*(r))}	
\end{eqnarray}
and for the left wedge it has the following form
\begin{eqnarray}
U&=&e^{\kappa^{(q)}(t+r^*(r))}\nonumber\\
V&=&-e^{-\kappa^{(q)}(t-r^*(r))}	
\end{eqnarray}
By using the metric in Kruskal coordinate, we can proceed to construct the Penrose diagram for a two-sided eternal QCSBH geometry. This we provide in Fig.\eqref{Fig1}. The horizon structure of the quantum-corrected Schwarzschild black hole is indeed very similar to that of the Reissner-Nordström (RN) black hole. However, it is important to note that while the RN black hole arises from adding a gauge field to the Einstein-Hilbert action, the black hole we have considered is obtained from the functional renormalization group approach. Further,  expanding eq.\eqref{lap} upto $\mathcal{O}(\tilde{\omega})$, the lapse function has the form
\begin{eqnarray}
f(r)\approx 1-\frac{2M G_{0}}{r}+\frac{2M \tilde{\omega}G_{0}^2}{r^3}~.
\end{eqnarray}
This is significantly different from the RN black hole. Hence, the physics for the QCSBH and the RN black hole are quite different.\\
We now proceed to explore the proposals given in \cite{Saha:2021ohr, RoyChowdhury:2022awr,RoyChowdhury:2023eol} for the black hole solution discussed above.
It is to be noted that qualitatively the QCSBH solution is an improved solution resulting from the flow of the Newton's gravitational constant $G$ arising from the functional renormalization group approach to quantum gravity. The flow equation for $G$ following from this approach does not introduce any higher curvature corrections to the Einstein-Hilbert action which can potentially modify the area term of the island formula (given in eq.\eqref{eq1}), as shown in \cite{Alishahiha_2021}. Some previous studies related to island formalism for modified theories of gravity can be found in \cite{Tong:2023nvi,Anand:2023ozw,Jain:2023xta,Omidi:2021opl,Yadav:2022fmo,Anand:2022mla,Yadav:2022mnv,Djordjevic:2022qdk,Liu:2025flo,Dong:2025duz,Alishahiha:2020qza,Lu:2021gmv}. 
 \\
\section{After Page time scenario and the role of mutual information of subsystems in the Page curve}\label{Sec2}
\begin{figure}[htb]
		\centering
		\includegraphics[scale=0.55]{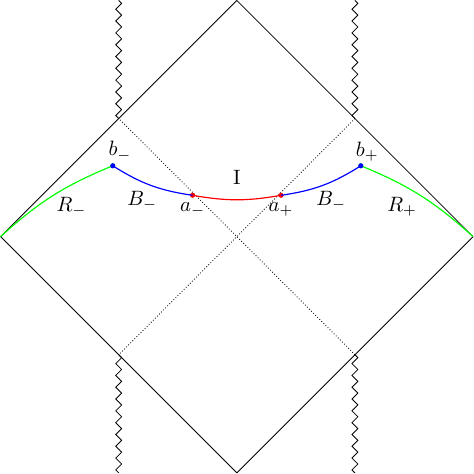}
		\caption{Penrose diagram specifying the island region (in red) with the boundaries $a_{\pm}=(\pm t_a,a)$ and the $B_{\pm}$ regions are denoted in blue. The inner boundaries of $B_{\pm}$ regions are $a_{\pm}=(\pm t_a,a)$ and the outer boundaries are $b_{\pm}=(\pm t_b,b)$.}
		\label{fig2}
	\end{figure}

\noindent As we have already discussed, the paradoxical situation arises just after the Page time when fine-grained entropy of Hawking radiation exceeds the thermal (coarse-grained) entropy of the black hole and the inclusion of the island contributions helps us to overcome this unease situation. In short, we need to make use of the formula given in eq.\eqref{eq1} in order to compute the correct form of fine-granied entropy of black hole radiation. 
\noindent The second term in eq.\eqref{eq1}, can be computed by considering the fact that, the quantum state on the full Cauchy slice is a pure state. This implies one can write down the following equality regarding the relevant von Neumann entropies
	\begin{equation}
	S_{vN}(I\cup R)=S_{vN}(I\cup R_+\cup R_-)=S_{vN}(B_+\cup B_-)~.
	\end{equation}
	\noindent The regions $B_{\pm}$ are are extended from the points $b_{\pm}=(\pm t_b,b)$ to the boundaries of the island region (location of the quantum extremal surfaces), that is, $a_{\pm}=(\pm t_a,a)$. This can be graphically represented with the help of a Penrose diagram, as given in Figure \eqref{fig2}. In order to compute the von Neumann entropy of the matter fields, we have considered that, the entire spacetime is filled with $2d$ free conformal matter (with large central charge $c$). 
 In the present scenario, our interest lies in the computation of the von Neumann entropy of matter fields on two disjoint subsystem, $S_{vN}(B_+ \cup B_-)$. This we can do by using the following formula \cite{Calabrese:2009ez}
\begin{widetext}
	\begin{eqnarray}\label{eq13}
	S_{vN}(B_+\cup B_-)=\left(\frac{c}{3}\right)\log\Big[\frac{d(a_+,a_-)d(b_+,b_-)d(a_+,b_+)d(a_-,b_-)}{d(a_+,b_-)d(a_-,b_+)}\Big]~.
	\end{eqnarray}
\end{widetext}
To get the explicit form of $S_{vN}(B_+ \cup B_-)$, we need to compute the individual distances given in the above formula by using the black hole metric given in eq.\eqref{Eq12}. Previous works in this direction suggested that for a generic spherically symmetric two-sided eternal black hole, one can write down the following \cite{Hashimoto:2020cas,Matsuo:2020ypv}
\begin{eqnarray}\label{eq14}
S_{vN}(B_+\cup B_-)&=&S_{vN}(B_+)+S_{vN}(B_-)\nonumber\\
&&+\sim\mathcal{O}(e^{-\frac{2\pi t_{a}}{\beta}}) +\sim\mathcal{O}(e^{-\frac{2\pi t_{b}}{\beta}})
\end{eqnarray}
where $S_{vN}(B_{\pm})$ has the form 
\begin{equation}\label{Eq36}
S_{vN}(B_{\pm})=\left(\frac{c}{3}\right)\log d(b_{\pm},a_{\pm})~.
\end{equation}
Furthermore, it was also suggested that one can ignore the time-dependent terms $\sim\mathcal{O}(e^{-\frac{2\pi t_{a}}{\beta}}), \mathcal{O}(e^{-\frac{2\pi t_{b}}{\beta}})$ in eq.\eqref{eq14} at the late times $(t_a,t_b\gg\beta)$. This consideration further simplifies $S_{vN}(B_+\cup B_-)$ in the following way
\begin{eqnarray}\label{eq15}
S_{vN}(B_+\cup B_-) \approx S_{vN}(B_+)+ S_{vN} (B_-).	
\end{eqnarray}
One then needs to substitute the above in eq.\eqref{eq1} and perform the extremization with respect to the island parameters $t_a$ and $a$, which results in $t_a\approx t_b$ and $a\approx r_+$. Now, substitution of these extremized values of island parameters, leads to the following form of fine-grained entropy of radiation after the Page time
\begin{equation}
S(R)=2S_{BH}+..~.
\end{equation}
The above result suggests that, there is no time-dependent piece in the expression of the $S(R)$, which is a desired feature for the eternal black holes as suggested by the Page curve. However, if we have a look into this analysis deeply, one can observe that in order to obtain the final time-independent expression of $S(R)$ one needs to ignore the exponentially time-dependent pieces, as shown in eq.\eqref{eq14} if we keep those terms in the expression of $S_{vN}(B_{+}\cup B_{-})$, one can explicitly check that it will not to be possible to obtain the mentioned result of $S(R)$. This observation is a very alarming one as one is basically neglecting time-dependent pieces in order to have a time-independent result!\\
In order to resolve this ambiguity, one can follow the approach originally given in \cite{Saha:2021ohr}. This approach can be understood in the following way. From the approximation given in eq.\eqref{eq15}, one can realize that it can be represented as
\begin{eqnarray}\label{mi1}
I(B_+:B_-)\equiv S_{vN}(B_+)+ S_{vN} (B_-)-S_{vN}(B_+\cup B_-)\approx 0\nonumber\\~.
\end{eqnarray}
The above equation is depicting the fact that one has to claim the mutual information between the matter fields on $B_{\pm}$ is approximately zero. Keeping this in mind, the following proposal was given in \cite{Saha:2021ohr} (and later followed in \cite{RoyChowdhury:2022awr,RoyChowdhury:2023eol}\\

\noindent\textbf{Proposal:}  \textit{Just after the Page time, when island starts to contribute significantly, the mutual information between the subsystems $B_+$ and $B_-$ vanishes excatly.}\\
	
\noindent This in turn means one has to demand the following
\begin{eqnarray}\label{eq16}
I(B_+:B_-)&=&0\nonumber\\
S_{vN}(B_+)+S_{vN}(B_-)&=&S_{vN}(B_+\cup B_-)~.	
\end{eqnarray} 

\noindent This proposal departs from the standard island prescription where we find that the mutual information between the regions $B_{+}$ and $B_{-}$ vanishes approximately (eq.\eqref{mi1}). Hence, in the standard island prescription, one requires extremization of the fine grained entropy with respect to the island parameters. As we shall find in the subsequent discussion, the vanishing of the mutual information between the regions $B_+$ and $B_-$ fixes the island parameter $t_{a}$.\\
To procced further, we make use of the eq(s).\eqref{eq13} and \eqref{Eq36} in the above which explicitly gives us
\begin{eqnarray}\label{neweq4}
d(a_{+},b_{-})d(a_{-},b_{+})&=&d(a_{+},a_{-})d(b_{+},b_{-})~.
\end{eqnarray}
Furthermore, if we substitute the above expression in eq.\eqref{eq13}, we get the following form for $S_{vN}(R_{+} \cup R_{-})$
\begin{eqnarray}\label{Eq41}
S_{vN}(B_{+}\cup B_{-})=\frac{2c}{3}\log d(a_{+},b_{+})~.
\end{eqnarray}
\begin{widetext}\label{neweq9}
\noindent The exact expression for various distances can be obtained from the black hole metric given in eq.\eqref{Eq12}. This reads
	\begin{eqnarray}
	d(a_+,b_+)&=&\sqrt{2F(a)F(b)e^{\kappa^{(q)}(r^{*}(b)+r^{*}(a))}}\Big[\cosh[\kappa^{(q)}(r^{*}(a)-r^{*}(b))]-\cosh[\kappa^{(q)}(t_{a}-t_{b})]\Big]^{\frac{1}{2}}=d(a_{-},b_{-})\nonumber\\
	d(a_-,b_+)&=&\sqrt{2F(a)F(b)e^{\kappa^{(q)}(r^{*}(b)+r^{*}(a))}}\Big[\cosh[\kappa^{(q)}(r^{*}(a)-r^{*}(b))]+\cosh[\kappa^{(q)}(t_{a}+t_{b})]\Big]^{\frac{1}{2}}=d(a_{+},b_{-})\nonumber\\
	d(b_{+},b_{-})&=&2F(b)e^{\kappa^{(q)} r^{*}(b)}\cosh(\kappa^{(q)} t_{b})\nonumber\\
	d(a_{+},a_{-})&=&2F(a)e^{\kappa^{(q)} r^{*}(a)}\cosh(\kappa^{(q)} t_{a})~.
	\end{eqnarray}
\end{widetext}
Interestingly, from the above expressions one can note that
\begin{eqnarray}
d(a_+,b_+)&=&d(a_-,b_-)\nonumber\\
d(a_+,b_-)&=&d(a_-,b_+)~.
\end{eqnarray}
Finally, we make use of these expression in eq.\eqref{neweq4} and obtain the following necessary condition for exact vanishing of mutual information (given in eq.\eqref{eq16})
\begin{eqnarray}\label{Eq49}
t_a-t_b=|r^*(a)-r^*(b)|~.
\end{eqnarray}
This expression helps us to represent one of the island parameters, that is, $t_a$ in terms of $t_b$ and $b$ and if we substitute that expression of $t_a$ in eq.\eqref{Eq41}, it gives us
\begin{eqnarray}\label{Eq48}
S_{vN}(B_+\cup B_-)=\frac{c}{3}\log\left[2F(a)F(b)e^{\kappa^{(q)}(r^{*}(b)+r^{*}(a))}\right].\nonumber\\	
\end{eqnarray}
Surprisingly, we have obtained a time-independent expression for $S_{vN}(B_+\cup B_-)$ without performing the conventional extremization of $t_a$. This is one of the key results of this work which we have obtained by using the proposal of exact vanishing of mutual information.\\
The next step is to find the current location quantum extremal sufaces. In order to do that we have to use the result obtained in eq.\eqref{Eq48} alongwith the area term, that is, $\frac{\mathrm{Area}(\partial I)}{4G_N}=2\times\frac{4\pi a^2}{4G_N}$ in eq.\eqref{eq1} and extremizing with respect to the island parameter $``a"$, that is, $\partial_a S(R)=0$. This in turn produces
\begin{eqnarray}\label{eq19}
a&=&r_+ - \left(\frac{cG_0}{24\pi}\right)\frac{1}{r_+}+...\nonumber\\
&\approx&a^{(S)}-\frac{\tilde{\omega}}{M}\left(\frac{cG_0}{48\pi}\frac{1}{r^{(S)}}+\frac{1}{2}\right)
\end{eqnarray}
where $a^{(S)}$ is the extremized value of $``a"$ which one can obtain for a usual classical Schwarzschild Schwarzshild black hole, given in \cite{Saha:2021ohr} \footnote{ The expression of $a_{(S)}$ given by, $a^{(S)}=r^{(S)}-\left(\frac{cG}{24\pi}\right)\frac{1}{r^{(S)}}$.}. It is to be mentioned that $r^{(S)}$ is the horizon radius of Schwarzschild black hole. The second term in the above result arises due to the inclusion of the quantum corrections to the classical geometry. The expression for $r_{+}$ has been provided in eq.\eqref{EQ8}. In the limit $\tilde{\omega}\rightarrow 0$, one can recover the result of island parameter for standard Schwarzschild black hole.\\
To obtain the final expression of the fine-grained entropy of Hawking radiation, we need to substitute the extremized value of ``$a$'' in eq.\eqref{Eq48}. This in turn yields
\begin{eqnarray}\label{EQ52}
S(R)&\approx&2S_{BH}^{(q)} -\frac{c}{6}\log\left(S_{BH}^{(q)}\right)-\frac{\left(\frac{c}{12}\right)^2}{2S_{BH}^{(q)}}+...~
\end{eqnarray}
where $S_{BH}^{(q)}$ is the coarse-grained entropy of QCSBH given in eq.\eqref{EQ13}.
It is to be mentioned that the above expression is free of any time dependency. The above expression also contains logarithmic and inverse power law correction terms, similar to our previous studies \cite{Saha:2021ohr,RoyChowdhury:2022awr,RoyChowdhury:2023eol}.
We can also rewrite the above result of $S(R)$ for QCSBH in terms of the same obtained for standard Schwarzschild blac hole, which is to be done by using eq.\eqref{EQ8}. This reads
\begin{eqnarray}
S(R)&\approx&S^{(S)}(R)+\tilde{\omega}\Bigg[\frac{\frac{c}{3}}{4G_{0}M^{2}}-\frac{S_{BH}^{(S)}}{M^2G_{0}}-\frac{\left(\frac{c}{12}\right)^2}{S_{BH}^{(S)}}\frac{1}{4M^2G_{0}}\Bigg]...\nonumber\\
&=&S^{(S)}(R)+\tilde{\omega}\pi\Bigg[\frac{\frac{c}{3}}{S_{BH}^{(S)}}-\frac{\left(\frac{c}{12}\right)^2}{(S_{BH}^{(S)})^2}-4\Bigg]
\end{eqnarray}
where $S^{(S)}(R)$ denotes the fine-grained entropy of Hawking radiation for the standard, classical Schwarzschild black hole solution  \cite{Saha:2021ohr}\footnote{The expression of $S^{(S)}(R)$ reads, $S^{(S)}(R)=2S^{(S)}_{BH}-\frac{c}{6}\log S^{(S)}_{BH}-\frac{(c/12)^2}{2S^{(S)}_{BH}}$.}.\\
On the other hand, by subsituting the extremized results of the island parameters in eq.\eqref{Eq49}, we obtain the following form of the condition of vanishing mutual information 
\begin{eqnarray}\label{eq21}
t_a-t_b= \left(\frac{\beta^{(q)}}{4\pi}\right)\log\left(S_{BH}^{(q)}\right)=t_{Scr}^{(q)}~
\end{eqnarray}
where $t_{Scr}^{(q)}$ is the \textit{scrambling time} \cite{Sekino:2008he,Hayden:2007cs} for the QCSBH spacetime. For the sake of completeness, we now express the scrambling time for QCSBH in the following way 
\begin{eqnarray}
t_{Scr}^{(q)}&\approx&t_{Scr}^{(S)}+\frac{\beta^{(S)}}{4\pi}\Bigg[\frac{\tilde{\omega}G_{0}}{(r^{(S)})^2}\log S_{BH}^{(S)}\nonumber\\&+&\left(1+\frac{\tilde{\omega}G_{0}}{(r^{(S)})^2}\right)\log\left(1-\frac{\tilde{\omega}}{Mr^{(S)}}\right)\Bigg]+...\nonumber\\
&=&t_{Scr}^{(S)}+\frac{\beta^{(S)}}{4\pi}\Bigg[\frac{\pi\tilde{\omega}}{S_{BH}^{(S)}}\log S_{BH}^{(S)}\nonumber\\&+&\left(1+\frac{\pi\tilde{\omega}}{S_{BH}^{(S)}}\right)\log\left(1-\frac{\tilde{2\pi\omega}}{S_{BH}^{(S)}}\right)\Bigg]
\end{eqnarray}
where $t_{Scr}^{(S)}$ is the scrambling time for the classical Schwarzschild black hole \footnote{The scrambling time for Schwarzschild black hole reads $t_{Scr}^{(S)}=\frac{\beta^{(S)}}{4\pi}\log S_{BH}^{(S)}$}. To get the the above result we have used the relations given in eq.\eqref{EQ8} and eq.\eqref{EQ13}. It is interesting to observe that the above result suggests $t_{Scr}^{(q)}-t_{Scr}^{(S)}>0$, this in turn implies that the information recovery time is larger for the QCSBH in comparison to the standard Schwarzschild black hole. The fact that the information recovery time is longer in case of the QCSBH refers to the fact that it takes more time to scramble the information in case of the QCSBH which arises from the asymptotic safe theory of quantum gravity. This is in line with the fact that in a quantum system, the scrambling time is longer. Hence, the entire process of recovering the information after it has been scrambled in the QCSBH is expected to be longer compared to the standard Schwarzschild black hole.\\
Now we will make few comments regarding the obtained result. Keeping in mind the relation between entanglement wedge and mutual information \cite{Takayanagi:2017knl}, one can argue that as long as $t_a-t_b< t_{Scr}^{(q)}$, the entanglement wedge associated with $B_{+}\cup B_{-}$ is in the connected phase as the mutual information between $B_{+}$ and $B_{-}$ is nonzero. However, as soon as the difference between $t_{a}$ and $t_{b}$, that is, $t_a-t_b$ equals $t_{Scr}^{(q)}$, the entanglement wedge associated with $B_{+}\cup B_{-}$ is in the disconnected phase as mutual information between $B_+$ and $B_-$ vanishes. \\
Finally, we will compute the Page time $(t_{P}^{(q)})$ for the QCSBH. This can be obtained by equating the leading order result of eq.\eqref{EQ24} and eq.\eqref{EQ52}. The expression of $(t_{P}^{(q)})$ reads
\begin{eqnarray}\label{EQ56}
t_P^{(q)} \approx \left(\frac{3\beta^{(q)}}{\pi c}\right) S_{BH}^{(q)}~.
\end{eqnarray}
Now using  eq.\eqref{EQ8} and eq.\eqref{EQ13} in the above result we obtain
\begin{eqnarray}
t_{P}^{(q)}&\approx& t_{P}^{(S)}\left(1-\frac{\tilde{\omega}G_{0}}{(r^{(S)})^2}\right)\nonumber\\
&=&t_{P}^{(S)}\left(1-\frac{\pi \tilde{\omega}}{S_{BH}^{(S)}}\right)
\end{eqnarray}
\noindent where $t_{P}^{(S)}$ is the Page time for the standard Schwarzschild black hole, the expression of $t_{P}^{(S)}$ reads $t_{P}^{(S)}=\frac{3\beta^S}{\pi c}S_{BH}^S$. As we have aleady mention that the island starts contribute to the von-Neumann entropy of Hawking radiation just after the Page time. Therefore, our result suggests that for a quantum-corrected Schwarzschild black hole, island formation begins earlier compared to the standard Schwarzschild black hole. The fact that the island formation begins earlier in case of the QCSBH compared to the standard Schwarzschild black hole indicates that the island formation needed for getting the correct time evolution of von-Neumann entropy of Hawking radiation is favoured by the QCSBH which arises from the asymptotic safe theory of quantum gravity. 
The reason behind this can be understood from the following observation. 
It is clear that quantum gravity plays an important role in the resolution of the 
information loss paradox as  the island formalism is an effective quantum gravity theory.  The QCSBH solution, on the other hand, is derived from an effective action of gravity and captures some elements of quantum gravity which in turn helps in the island formation process earlier needed to get the correct time evolution of the von Neumann entropy of Hawking radiation. \\
We would like to make few comments now. In this section, we have shown that inclusion of the island contribution in turn leads us to the exact vanishing of the mutual information between the matter fields localized on the subsystems $B_+$ and $B_-$ which eventually yields a time-independent expression for the fine-grained entropy of Hawking radiation. Keeping this in mind, one can may rewrite the formula given in \eqref{eq1} in the following form
\begin{eqnarray}\label{island_new}
S(R)&=&\textrm{min}~\mathop{\textrm{ext}}_{\mathcal{\mathrm{I}}}\bigg\{\frac{\textrm{Area}(\partial I)}{4G_N}+S_{vN}(B_+ \cup B_-)\Big\vert_{I(B_+:B_-)=0}\bigg\}\nonumber\\
&=&\textrm{min}~\mathop{\textrm{ext}}_{\mathcal{\mathrm{I}}}\bigg\{\frac{\textrm{Area}(\partial I)}{4G_N}+S_{vN}(B_+)+ S_{vN}(B_-)\bigg\}~.\nonumber\\
\end{eqnarray}
In the first line we make use of the fact that the state on the full Cauchy slice is a pure state and based upon this observation one can write $S_{vN}(B_+ \cup B_-)=S_{vN}(I \cup R)$. Furthermore, it is subject to the condition of vanishing of mutual information between $B_+$ and $B_-$. In the final form of the formula we have incorporated the the fact $t_a-t_b=|r^*(a)-r^*(b)|$ which in turn fixes one of the island parameters, namely, $t_a$ and only $``a"$ is left to be extremized. We propose the above formula for evaluating the fine-grained entropy of Hawking radiation for an eternal black hole.
\section{Before Page time scenario and the emergence of Hartman-Maldacena time}\label{Sec3}
\begin{figure}[htb]
 \centering
 	\includegraphics[scale=0.55]{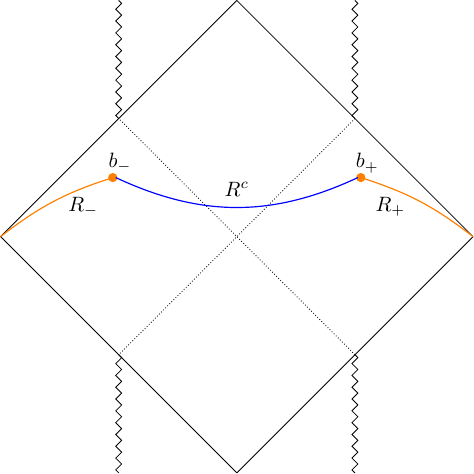}
 \caption{The $R_{\pm}$ regions have been shown in orange with the inner boundaries $b_{\pm}=(\pm t_b,b)$ and the complementary region $R^c$ is in blue.}
 \label{fig1}
 \end{figure}

\noindent In this time domain, that is before Page time ($t_{b}\le t_{P}$), the contribution of island is insignificant. Therefore the fine-grained entropy of Hawking radiation is identified with the von-Neumann entropy of the matter fields on $R_{+}\cup R_{-}$,  where the $\pm$ indicates the locations of the regions in right and the left wedge of the Penrose diagram respectively. This can be written down as
\begin{equation}
S(R)=S_{vN}(R_{+}\cup R_{-})~.
\end{equation}
Further, by using the property of the pure state one can argue
\begin{eqnarray}\label{Eq8}
S_{vN}(R_+\cup R_-)&=&S_{vN}(R^c)	
\end{eqnarray}
where $R^{c}$ is the complementary region of $R_{+}\cup R_{-}$. The two end points of $R_{\pm}$ are $[e_{\pm}:b_{\pm}]$.
 As $R_{\pm}$ regions are extended to spatial infinity from the inner boundary $b_{\pm}=(\pm t_b,b)$, we introduce the point $e_{\pm}$ in order to regularize it, that is, $e_{\pm}=(0,e)$ \cite{RoyChowdhury:2022awr}. All the results are computed in the limit $e\rightarrow\infty$.
 This has been shown graphically in the Penrose diagram given in Fig.(\ref{fig1}).
Now, similar to the analysis we have shown before, the von Neumann entropy of matter fields in this case reads \cite{Calabrese:2009qy}
\begin{eqnarray}\label{sr}
S_{vN}(R_{+}\cup R_{-})=S_{vN}(R^c)=\left(\frac{c}{3}\right)\log d(b_+,b_-)~.	
\end{eqnarray}
The distance $d(b_+,b_-)$ can be computed from the metric given in eq.\eqref{Eq12}. This reads
\begin{eqnarray}
	d(b_{+},b_{-})=2F(b) e^{\kappa^{(q)} r^{*}(b)}\cosh(\kappa^{(q)} t_{b})~.
\end{eqnarray}
By using the above expression it is quite easy to show the following result for the fine-grained entropy 
\begin{eqnarray}\label{Neweq1}
	S(R)&=&S_{vN}(R_+\cup R_-)\nonumber\\
 &=&\left(\frac{c}{3}\right)\log\left[\left(\frac{\beta^q}{\pi}\right) \sqrt{f(b)}\cosh\left(\frac{2\pi t_b}{\beta^q}\right)\right]~.	
\end{eqnarray}
In the above result $\beta^q$ is the inverse of Hawking temperature for the QCSBH black hole.\\
Now we will study the behaviour of $S(R)$ in two different time domains, namely, in the early time domain $t_b\ll\beta^{(q)}$ and in the late time domain $t_{P}>t_b\gg\beta^{(q)}$. In the early time domain it has the following form
\begin{eqnarray}\label{Eq21}
S(R) &\approx& \left(\frac{c}{3}\right)\log\left[\left(\frac{\beta^{(q)}}{\pi}\right) \sqrt{f(b)}\right]+\left(\frac{c}{6}\right)\left(\frac{2\pi t_b}{\beta^{(q)}}\right)^2~~.\nonumber\\	
\end{eqnarray}
One can obtain the above result from eq.\eqref{Neweq1} by incorporating the fact $t_b\ll\beta^{(q)}$. It is observed that, in this time domain the EE of radiation increases quadratically with the observer's time.\\
On the other hand, in the late time ($t_{P}>t_b\gg\beta^{(q)}$) it reads
\begin{eqnarray}\label{EQ24}
S(R) &\approx&	\left(\frac{c}{3}\right)\log\left[\left(\frac{\beta^{(q)}}{\pi}\right) \sqrt{f(b)}\right]+\left(\frac{c}{3}\right)\left(\frac{2\pi t_b}{\beta^{(q)}}\right)~~.\nonumber\\
\end{eqnarray}
From the above given expressions one can note that in the early time domain $S(R)$ has quadratic time dependence, that is $S(R)\sim t_{b}^2$. On the other hand in the late time domain, $S(R)$ grows linearly over observer's time, $S(R)\sim t_{b}$ which matches with Hawking's suggestion. This observation is agreed with \cite{Hartman:2013qma}.\\
In this section our aim is to compute the mutual information between the matter fields localized on region $R_+$ and $R_-$. The reason behind this interest is the following. In the last section, we have already seen the importance of mutual information in obtaining the Page curve. In the before Page time scenario, although there is no island contribution but it will be very interesting to see how does the mutual information evolves over time. To compute the von Neumann entropy of matter fields on $R_{\pm}$, we will use the formula given below \cite{Calabrese:2009qy}
\begin{eqnarray}\label{sr+}
	S_{vN}(R_+)&=&\left(\frac{c}{3}\right)\log d(b_+,e_+)\nonumber\\
	S_{vN}(R_-)&=&\left(\frac{c}{3}\right)\log d(b_-,e_-)~.
\end{eqnarray}
The relevant distances can be computed by using the black hole metric given in eq.\eqref{Eq12}. This reads
\begin{eqnarray}
d(b_{\pm},e_{\pm})
	&=&\sqrt{2F(b)F(e)e^{\kappa^{(q)} r^{*}(b)}[\cosh(\kappa^{(q)} r^{*}(b))-\cosh(\kappa^{(q)} t_{b})]}~.\nonumber\\
\end{eqnarray}
The above result was obtained by using the fact that, in the limit $e\rightarrow \infty$, $r^{*}$ vanishes. Now, using the above results of $d(b_{\pm},e_{\pm})$ in eq.\eqref{sr+} we get
\begin{widetext}
\begin{eqnarray}\label{Neweq2}
S_{vN}(R_+) = S_{vN}(R_-) =\left(\frac{c}{6}\right)	\log\left[2\left(\frac{\beta^{(q)}}{2\pi}\right)^2\sqrt{f(b)f(e)}\left\{|\cosh\left(\frac{2\pi r^*(b)}{\beta^{(q)}}\right)-\cosh\left(\frac{2\pi t_b}{\beta^{(q)}}\right)|\right\}\right]~.	
\end{eqnarray}
\end{widetext}
\noindent With these above computed necessary results in hand, we can proceed to compute the mutual information between $R_{+}$ and $R_{-}$, that is $I(R_{+}\cup R_{-})$. This yields
\begin{widetext}
 	\begin{eqnarray}\label{Eq26}
 	I(R_+:R_-) &=& S_{vN}(R_+)+ S_{vN}(R_-)-S_{vN}(R_+\cup R_-)\nonumber\\
 	&=&\left(\frac{c}{3}\right)\log\left[\left(\frac{\beta^{(q)}}{2\pi}\right)\sqrt{f(e)}\left\{\frac{|\cosh\left(\frac{2\pi r^*(b)}{\beta^{(q)}}\right)-\cosh\left(\frac{2\pi t_b}{\beta^{(q)}}\right)|}{\cosh\left(\frac{2\pi t_b}{\beta^{(q)}}\right)}\right\}\right]~.\nonumber\\
 	\end{eqnarray}
\end{widetext}
For the sake of better understanding, we look at the behaviour of $I(R_{+}\cup R_{-})$ in two different time domains, namely in the early time and in the late time domain. In the early time domain, that is when $t_{b}\ll \beta^{(q)}$, the expression of the mutual information between $R_{+}$ and $R_{-}$ reduces to the following form
\begin{widetext}
\begin{eqnarray}\label{Eq27}
	I(R_+:R_-)&\approx&\left(\frac{c}{3}\right)\left[\log\left[\left(\frac{\beta^{(q)}}{2\pi}\right)\sqrt{f(e)}\cosh\left(\frac{2\pi r^*(b)}{\beta^{(q)}}\right)\right]-\mathrm{sech}\left(\frac{2\pi r^*(b)}{\beta^{(q)}}\right)-\left(\frac{2\pi^2}{\beta^{(q)^2}}\right)\left\{1+\mathrm{sech}\left(\frac{2\pi r^*(b)}{\beta^{(q)}}\right)\right\}t_b^2\right]~.\nonumber\\
\end{eqnarray}
\end{widetext}
The above expression suggests that at the very begining, that is, at $t_{b}=0$, $I(R_+:R_-)$ is non-zero and finite. It then decreases with respect to the observer's time $t_b$. On the other hand at late time, that is for $t_{b}\gg\beta^{(q)}$, $I(R_+:R_-)$ boils down to the following form
\begin{widetext}
\begin{eqnarray}\label{Eq28}
I(R_+:R_-)&\approx&\left(\frac{c}{3}\right)\left[\log\left[\left(\frac{\beta^{(q)}}{2\pi}\right)\sqrt{f(e)}\right]-2\cosh\left(\frac{2\pi r^*(b)}{\beta^{(q)}}\right)e^{-\left(\frac{2\pi t_b}{\beta^{(q)}}\right)}\right]~.	
\end{eqnarray}
\end{widetext}
The above result of $I(R_+:R_-)$ implies that at late time $I(R_+:R_-)$ increases with respect to $t_b$. This implies the behaviour of $I(R_+:R_-)$ in two different time domains are different, in particular, at the early time $I(R_+:R_-)$ is a decreasing function with respect to time and in the late time domain it is increasing. This in turn means that there is a lowest value of $I(R_+:R_-)$ which characterizes this change in its nature. In order to have a concrete idea about the possible lowest value, we plot the general expression of $I(R_+:R_-)$ (eq.\eqref{Eq26}) with respect to the observer's time $t_b$. In Fig.\eqref{f1} we have given this where the vertical axis represents $\mathcal{I}_R=\left(\frac{3}{c}\right)I(R_+:R_-)$ and the horizontal axis depicts $t_b$. Further, we have set $r^{(S)}=1$, $b=2r^{(S)}=2$ and $G_0=1$. We have also taken the limit $e\rightarrow\infty$. We observe that irrespective of the value for the quantum-correction paramater $\tilde{\omega }$, the mutual information between $R_+$ and $R_-$ vanishes for a particular value of the observer's time $t_b$. This observation motivates us to look for a analytical expression corresponding to this particular time scale, namely, $t_{R}$.
\begin{widetext}
	~\\ 
 \begin{figure}[htb]
	\centering
	\includegraphics[scale=0.45]{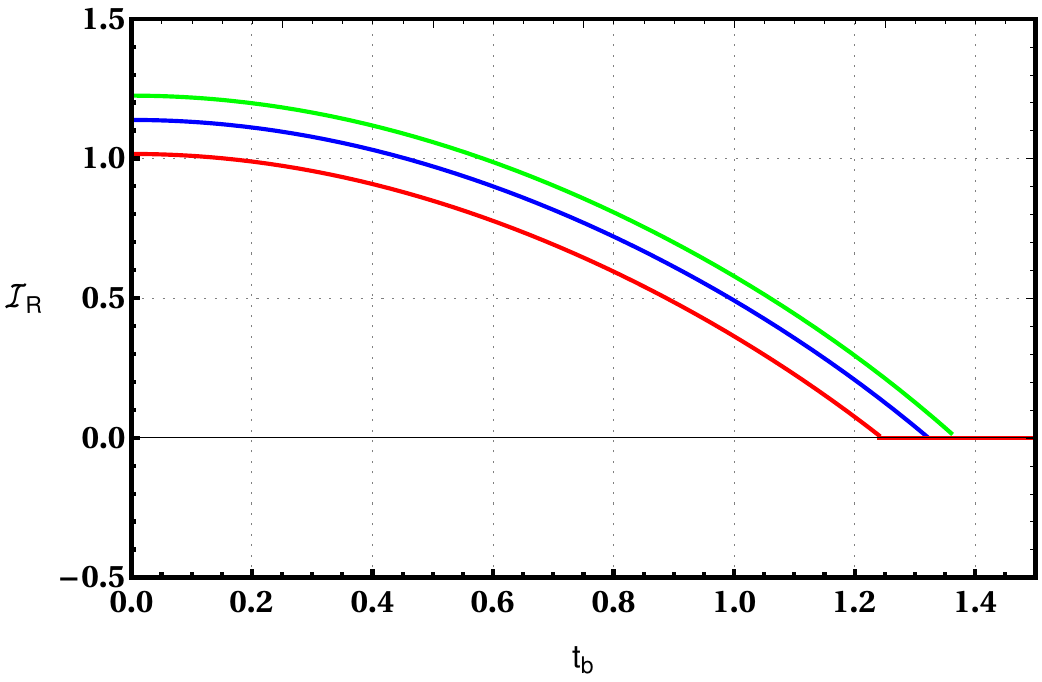}
	\caption{The figure shows the variation of $\mathcal{I}_R$ with respect to the observer's time $t_b$ for different values of \( \tilde{\omega} \). The red, blue, and green curves represent the results for \( \tilde{\omega} = 0, 0.1, 0.2 \) respectively. Here, we have set $r^{(S)}=1, b=2r^{(S)}=2$ and $G_{0}=1$.}
	\label{f1}
\end{figure}
\end{widetext}
The vanishing of the mutual information at $t_{b}=t_{R}$  means that the entanglement wedge associated with $R_{+}\cup R_{-}$ become disconnected. This was first discussed in \cite{RoyChowdhury:2022awr,RoyChowdhury:2023eol} with the help of the following proposal\\ 
	
\begin{widetext}
	\noindent\textbf{Proposal:} \textit{Starting from a finite, non-zero value (at $t_b=0$), the mutual information between the subsyems $R_+$ and $R_-$ vanishes at a particular value of the observer's time ($t_b=t_R$).}\\
	
\noindent In order to find this time-scale $t_{R}$, we need to solve the following equation
\begin{eqnarray}\label{eq7}
	I(R_+:R_-)|_{t_b=t_R}&=&0~.
\end{eqnarray}
This in turn yields the following form of $t_{R}$
\begin{eqnarray}
	t_R=&&\left(\frac{\beta^{(q)}}{2\pi}\right)\cosh^{-1}\left\{\left(\frac{\frac{\beta^{(q)}}{2\pi}\sqrt{f(e)}}{1+\frac{\beta^{(q)}}{2\pi}\sqrt{f(e)}}\right)\cosh\left(\frac{2\pi r^*(b)}{\beta^{(q)}}\right)\right\}.\nonumber\\
\end{eqnarray}
\noindent We would like to make few comments regarding the obtained result. Firstly, as we have have menioned earlier, the importance of $t_R$ lies in the fact that the mutual information between the matter fields on $R_{+}$ and $R_{-}$ vanishes eaxctly. Another thing to be noted that this time-scale resides in the early time domain, that is, $t_{R}< \beta^{(q)}$. Now we will compute $S(R_{+}\cup R_{-})$ at $t_{b}=t_{R}$. The expression of fine-grained entropy of Hawking radiation at this particular moment of time reads

\begin{eqnarray}\label{eq12}
S_{vN}^{t_{b}=t_{R}}(R_+\cup R_-)&=&\frac{c}{3}\log\left[\frac{\left(\frac{\beta^{(q)}\sqrt{f(e)}}{2\pi}\right)^{2}}{1+\frac{\beta^{(q)}\sqrt{f(e)}}{2\pi}}\cosh\left(\frac{2\pi r^{*}(b)}{\beta^{(q)}}\right)\right]\nonumber\\
&\approx&\frac{c}{3}\log\left[\frac{\beta^{(q)}}{2\pi}\sqrt{f(e)}\right]+\frac{c}{6}\left(\frac{r_{+}}{b}\right)^{2}~.\nonumber\\	
\end{eqnarray}
\end{widetext}
So, our proposal suggests that in the time domain $0\leq t_b<t_R$, the mutual information between the matter fields on $R_{+}$ and $R_{-}$ is non zero and the entanglement wedge associated with $R_{+}\cup R_{-}$ is in its connected pahse. However, at $t_{b}=t_{R}$ the mutual correlation vanishes ($I(R_{+}\cup R_{-})=0$), this means that at $t_{b}=t_{R}$ the entanglement wedge corresponding to $R_{+}\cup R_{-}$ is in the disconnected pahse. These observations strongly indicate that this time-scale $t_R$ is nothing but the Hartman-Maldacena time $t_{HM}$ \cite{Hartman:2013qma}. At the particular time-scale $t_b=t_R$, the entanglement surface corresponding to $R_+\cup R_-$ splits and leads to $S_{vN}(R_+\cup R_-)=S_{vN}(R_+)+ S_{vN}(R_-)$ which results in a vanishing mutual information between $R_+$ and $R_-$. This particular feature of $t_R$ suggests that it is precisely the Hartman-Maldacena time corresponding to the time-evolution of mutual information $I(R_+:R_-)$ at which it saturates to the value $I(R_+:R_-)=0$. This observation is in line with the one given in \cite{Grimaldi:2022suv}.
 Further, the expression of $I(R_+:R_-)$ at $t_b=\beta^{(q)}$ (by substituting $t_b =\beta^{(q)}$ in eq.\eqref{Eq26}), is given by
\begin{eqnarray}
I(R_+:R_-) &=&\left(\frac{c}{3}\right)\log\left[\left(\frac{\beta^{(q)}}{2\pi}\right)\sqrt{f(e)}\cosh\left(\frac{2\pi r^*(b)}{\beta^{(q)}}\right)\right]~.\nonumber\\
\end{eqnarray}
The above result suggests that after the Hartman-Maldacena time, the mutual information between the matter fields on $R_+$ and $R_-$ begins to increase.
\section{Conclusion}\label{Sec4}
\noindent We will now summerize our findings. In this paper, we have investigated the crucial role played by the mutual information (in the after Page time scenario) between the subsystems (relevant regions where the matter fields are localized) $B_+$ and $B_-$ in obtaining the correct time-evolution of fine-grained entropy of Hawking radiation, as suggested by Page. This was done by utilizing a proposal first given in \cite{Saha:2021ohr} and later followed in \cite{RoyChowdhury:2022awr,RoyChowdhury:2023eol}. This we have done for a quantum corrected or more precisely, a functional renormalization group improved two-sided eternal Schwarzschild black hole \cite{Bonanno:2000ep}. The motivation for considering such non-trivial geometry lies in the fact that this solution fimrly represents the essence of an ``effective" quantum theory of gravity as it arises from asymptotically safe effective average action $\Gamma_{k}[g_{\mu\nu}]$ which can correctly describe all gravitational phenomena, including the effects of all loops \cite{Reuter:1996cp}. Further, the mentioned action leads to a running nature for the usual Newton's gravitational constant which discriminates this geometry from the standard Schwarzschild black hole solution. On the other hand, the mentioned proposal regarding mutual information can be justified form the following facts. One can note that just after the Page time $t_P$, replica wormhole saddle point dominates and the island starts to contribute significantly and one has to make use of the formula given in eq.\eqref{eq1}, in order to compute the fine-grained entropy of Hawking radiation correctly where the first term provides us a geometric contribution and the second term quantifies the matter field contribution. Further, based upon the pure state property one can write down $S_{vN}(I\cup R)=S_{vN}(I\cup R_{+}\cup R_{-})=S_{vN}(B_{+}\cup B_{-})$. In earlier works in this direction it was observed that by using 
the late time approximated result of $S_{vN}(B_{+}\cup B_{-})$, that is,  $S_{vN}(B_{+}\cup B_{-})\sim S_{vN}(B_{+})+S_{vN}(B_{-})$ along with the extremized values of the island parameters ($t_{a}\sim t_{b}$ and $a\sim r_{H}$), one gets the desired time-independent result for $S(R)$ corresponding to a two-sided eternal black hole geometry in the after Page time scenario. However, the approximation result of $S_{vN}(B_{+}\cup B_{-})\sim S_{vN}(B_{+})+S_{vN}(B_{+})$ deals with the fact that one has to ignore time-dependent terms $\sim e^{-\frac{2\pi t_{b,a}}{\beta^q}}$ and if not neglected, obtaining a time-independent expression for $S(R)$ would not be possible. This in turn means one has to ignore certain time-dependent terms in order to have a desired time-independent result. We resolved this ambiguity in \cite{Saha:2021ohr,RoyChowdhury:2022awr,RoyChowdhury:2023eol} by demanding that the mutual information between $B_{+}$ and $B_{-}$ vanishes exactly, purely due to the emergence of island just after the Page time. We follow the same path here. This justified demand of vanishing mutual correlation leads to the associated condition $t_{a}-t_{b}=|r^{*}(a)-r^{*}(b)|$ and by utilizing this condition we have finally obtained the time-independent expression of $S_{vN}(I\cup R)$. This is a very important result as it helps one to overcome the mentioned ambiguity of neglecting certain time-dependent terms and also one obtains the needed value of $t_a$ without performing a extremization. Now extremizing the result of $S(R)$ with respect to the island parameter ``$a$", we have found out the position of the quantum extremal surfaces. Based upon this observations, we also propose a modified form of the island formula (given in \eqref{island_new}) which one may use to evaluate the fine-grained entropy of Hawking radiation for an eternal black hole. We make use of this extremized value of $``a"$ in two different places, firstly, by using it we get the final expression for $S(R)$ and on the other hand it simplifies the condition for vanishing of mutual information in the following way $t_{a}-t_{b}\sim t_{Scr}^q$. This result reflects the fact that as long as $t_{a}-t_{b}< t_{Scr}^q$, the mutual information is non-zero and the entanglement wedge of $B_{+}\cup B_{-}$ is in its connected phase and for $t_{a}-t_{b}= t_{Scr}^q$, the mutual information vanishes, depicting the disconnected phase of entnaglement wedge associated to $B_{+}\cup B_{-}$. Another interseting fact is that, the scrambling time for the quantum corrected Schwarzschild solution is greater than the same for standard Schwarzschild geometry. This in turn means that the information recovery time for the quantum corrected black hole is larger than the usual Schwarzschild black hole. We have also obtained the expression of the Page time for the mentioned geometry. We have found that the Page time for the quantum corrected Schwarzschild black hole is smaller compared to that of Schwarzschild black hole. Based upon this, one may say that island formation starts earlier for the QCSBH in comparison to the Schwarzschild black hole. Finally, we also study the time-dependent behaviour of relevant mutual information, that is, $I(R_+:R_-)$ in both the early and late time domain. Further, this leads us to the observation that during the early time $I(R_+:R_-)$ decreases with respect to the observer's time and at late time it increases with time. This gives us a hint that, there must exists a value of the observer's time at which the mutual information between $R_{+}$ and $R_{-}$ vanishes and the entanglement wedge associated to $R_{+}\cup R_{-}$ gets in its disconnected phase. We compute the explicit expression for the mentioned time-scale and observe that $S(R)|_{t_b=t_R}\sim\log\beta$. This observation leads us to identify the time $t_{R}$ as the Hartman-Maldacena time.

\section{Acknowledgements}
\noindent AS would like to thank S.N. Bose National Centre for Basic Sciences for the financial support through its Advanced Postdoctoral Research Programme. ARC would like to thank the organizers of \textit{Strings 2025} held at \textit{NYU Abu Dhabi}, as some of the results related to this work were presented there. The authors would also like to thank the organizers of \textit{XXVI DAE-BRNS High Energy Physics Symposium 2024} held at \textit{Banaras Hindu University, Varanasi}, as some of the results related to this work were presented there. The authors would like to thank the referees for very useful and crucial comments.
\bibliography{Reference.bib}

\begin{thebibliography}{122}%
\makeatletter
\providecommand \@ifxundefined [1]{%
 \@ifx{#1\undefined}
}%
\providecommand \@ifnum [1]{%
 \ifnum #1\expandafter \@firstoftwo
 \else \expandafter \@secondoftwo
 \fi
}%
\providecommand \@ifx [1]{%
 \ifx #1\expandafter \@firstoftwo
 \else \expandafter \@secondoftwo
 \fi
}%
\providecommand \natexlab [1]{#1}%
\providecommand \enquote  [1]{``#1''}%
\providecommand \bibnamefont  [1]{#1}%
\providecommand \bibfnamefont [1]{#1}%
\providecommand \citenamefont [1]{#1}%
\providecommand \href@noop [0]{\@secondoftwo}%
\providecommand \href [0]{\begingroup \@sanitize@url \@href}%
\providecommand \@href[1]{\@@startlink{#1}\@@href}%
\providecommand \@@href[1]{\endgroup#1\@@endlink}%
\providecommand \@sanitize@url [0]{\catcode `\\12\catcode `\$12\catcode
  `\&12\catcode `\#12\catcode `\^12\catcode `\_12\catcode `\%12\relax}%
\providecommand \@@startlink[1]{}%
\providecommand \@@endlink[0]{}%
\providecommand \url  [0]{\begingroup\@sanitize@url \@url }%
\providecommand \@url [1]{\endgroup\@href {#1}{\urlprefix }}%
\providecommand \urlprefix  [0]{URL }%
\providecommand \Eprint [0]{\href }%
\providecommand \doibase [0]{http://dx.doi.org/}%
\providecommand \selectlanguage [0]{\@gobble}%
\providecommand \bibinfo  [0]{\@secondoftwo}%
\providecommand \bibfield  [0]{\@secondoftwo}%
\providecommand \translation [1]{[#1]}%
\providecommand \BibitemOpen [0]{}%
\providecommand \bibitemStop [0]{}%
\providecommand \bibitemNoStop [0]{.\EOS\space}%
\providecommand \EOS [0]{\spacefactor3000\relax}%
\providecommand \BibitemShut  [1]{\csname bibitem#1\endcsname}%
\let\auto@bib@innerbib\@empty
\bibitem [{\citenamefont {Hawking}(1974)}]{Hawking:1974rv}%
  \BibitemOpen
  \bibfield  {author} {\bibinfo {author} {\bibfnamefont {S.~W.}\ \bibnamefont
  {Hawking}},\ }\bibfield  {title} {\enquote {\bibinfo {title} {{Black hole
  explosions}},}\ }\href {\doibase 10.1038/248030a0} {\bibfield  {journal}
  {\bibinfo  {journal} {Nature}\ }\textbf {\bibinfo {volume} {248}},\ \bibinfo
  {pages} {30--31} (\bibinfo {year} {1974})}\BibitemShut {NoStop}%
\bibitem [{\citenamefont {Hawking}(1975)}]{Hawking:1975vcx}%
  \BibitemOpen
  \bibfield  {author} {\bibinfo {author} {\bibfnamefont {S.~W.}\ \bibnamefont
  {Hawking}},\ }\bibfield  {title} {\enquote {\bibinfo {title} {{Particle
  Creation by Black Holes}},}\ }\href {\doibase 10.1007/BF02345020} {\bibfield
  {journal} {\bibinfo  {journal} {Commun. Math. Phys.}\ }\textbf {\bibinfo
  {volume} {43}},\ \bibinfo {pages} {199--220} (\bibinfo {year} {1975})},\
  \bibinfo {note} {[Erratum: Commun.Math.Phys. 46, 206 (1976)]}\BibitemShut
  {NoStop}%
\bibitem [{\citenamefont {Bekenstein}(1973)}]{Bekenstein:1973ur}%
  \BibitemOpen
  \bibfield  {author} {\bibinfo {author} {\bibfnamefont {Jacob~D.}\
  \bibnamefont {Bekenstein}},\ }\bibfield  {title} {\enquote {\bibinfo {title}
  {{Black holes and entropy}},}\ }\href {\doibase 10.1103/PhysRevD.7.2333}
  {\bibfield  {journal} {\bibinfo  {journal} {Phys. Rev. D}\ }\textbf {\bibinfo
  {volume} {7}},\ \bibinfo {pages} {2333--2346} (\bibinfo {year}
  {1973})}\BibitemShut {NoStop}%
\bibitem [{\citenamefont {Bekenstein}(1972)}]{Bekenstein:1972tm}%
  \BibitemOpen
  \bibfield  {author} {\bibinfo {author} {\bibfnamefont {J.~D.}\ \bibnamefont
  {Bekenstein}},\ }\bibfield  {title} {\enquote {\bibinfo {title} {{Black holes
  and the second law}},}\ }\href {\doibase 10.1007/BF02757029} {\bibfield
  {journal} {\bibinfo  {journal} {Lett. Nuovo Cim.}\ }\textbf {\bibinfo
  {volume} {4}},\ \bibinfo {pages} {737--740} (\bibinfo {year}
  {1972})}\BibitemShut {NoStop}%
\bibitem [{\citenamefont {Bekenstein}(1974)}]{PhysRevD.9.3292}%
  \BibitemOpen
  \bibfield  {author} {\bibinfo {author} {\bibfnamefont {Jacob~D.}\
  \bibnamefont {Bekenstein}},\ }\bibfield  {title} {\enquote {\bibinfo {title}
  {Generalized second law of thermodynamics in black-hole physics},}\ }\href
  {\doibase 10.1103/PhysRevD.9.3292} {\bibfield  {journal} {\bibinfo  {journal}
  {Phys. Rev. D}\ }\textbf {\bibinfo {volume} {9}},\ \bibinfo {pages}
  {3292--3300} (\bibinfo {year} {1974})}\BibitemShut {NoStop}%
\bibitem [{\citenamefont {Bardeen}\ \emph {et~al.}(1973)\citenamefont
  {Bardeen}, \citenamefont {Carter},\ and\ \citenamefont
  {Hawking}}]{Bardeen:1973gs}%
  \BibitemOpen
  \bibfield  {author} {\bibinfo {author} {\bibfnamefont {James~M.}\
  \bibnamefont {Bardeen}}, \bibinfo {author} {\bibfnamefont {B.}~\bibnamefont
  {Carter}}, \ and\ \bibinfo {author} {\bibfnamefont {S.~W.}\ \bibnamefont
  {Hawking}},\ }\bibfield  {title} {\enquote {\bibinfo {title} {{The Four laws
  of black hole mechanics}},}\ }\href {\doibase 10.1007/BF01645742} {\bibfield
  {journal} {\bibinfo  {journal} {Commun. Math. Phys.}\ }\textbf {\bibinfo
  {volume} {31}},\ \bibinfo {pages} {161--170} (\bibinfo {year}
  {1973})}\BibitemShut {NoStop}%
\bibitem [{\citenamefont {Wald}(2001)}]{Wald:1999vt}%
  \BibitemOpen
  \bibfield  {author} {\bibinfo {author} {\bibfnamefont {Robert~M.}\
  \bibnamefont {Wald}},\ }\bibfield  {title} {\enquote {\bibinfo {title} {{The
  thermodynamics of black holes}},}\ }\href {\doibase 10.12942/lrr-2001-6}
  {\bibfield  {journal} {\bibinfo  {journal} {Living Rev. Rel.}\ }\textbf
  {\bibinfo {volume} {4}},\ \bibinfo {pages} {6} (\bibinfo {year} {2001})},\
  \Eprint {http://arxiv.org/abs/gr-qc/9912119} {arXiv:gr-qc/9912119}
  \BibitemShut {NoStop}%
\bibitem [{\citenamefont {Nielsen}\ and\ \citenamefont
  {Chuang}(2000)}]{Chuang:2000}%
  \BibitemOpen
  \bibfield  {author} {\bibinfo {author} {\bibfnamefont {Michael~A.}\
  \bibnamefont {Nielsen}}\ and\ \bibinfo {author} {\bibfnamefont {Isaac~L.}\
  \bibnamefont {Chuang}},\ }\href@noop {} {\emph {\bibinfo {title} {Quantum
  Computation and Quantum Information}}}\ (\bibinfo  {publisher} {Cambridge
  University Press},\ \bibinfo {year} {2000})\BibitemShut {NoStop}%
\bibitem [{\citenamefont {Hawking}(1976)}]{PhysRevD.14.2460}%
  \BibitemOpen
  \bibfield  {author} {\bibinfo {author} {\bibfnamefont {S.~W.}\ \bibnamefont
  {Hawking}},\ }\bibfield  {title} {\enquote {\bibinfo {title} {Breakdown of
  predictability in gravitational collapse},}\ }\href {\doibase
  10.1103/PhysRevD.14.2460} {\bibfield  {journal} {\bibinfo  {journal} {Phys.
  Rev. D}\ }\textbf {\bibinfo {volume} {14}},\ \bibinfo {pages} {2460--2473}
  (\bibinfo {year} {1976})}\BibitemShut {NoStop}%
\bibitem [{\citenamefont {Page}(1993)}]{PhysRevLett.71.3743}%
  \BibitemOpen
  \bibfield  {author} {\bibinfo {author} {\bibfnamefont {Don~N.}\ \bibnamefont
  {Page}},\ }\bibfield  {title} {\enquote {\bibinfo {title} {Information in
  black hole radiation},}\ }\href {\doibase 10.1103/PhysRevLett.71.3743}
  {\bibfield  {journal} {\bibinfo  {journal} {Phys. Rev. Lett.}\ }\textbf
  {\bibinfo {volume} {71}},\ \bibinfo {pages} {3743--3746} (\bibinfo {year}
  {1993})}\BibitemShut {NoStop}%
\bibitem [{\citenamefont {Page}(2013)}]{Page_2013}%
  \BibitemOpen
  \bibfield  {author} {\bibinfo {author} {\bibfnamefont {Don~N}\ \bibnamefont
  {Page}},\ }\bibfield  {title} {\enquote {\bibinfo {title} {Time dependence of
  hawking radiation entropy},}\ }\href {\doibase 10.1088/1475-7516/2013/09/028}
  {\bibfield  {journal} {\bibinfo  {journal} {Journal of Cosmology and
  Astroparticle Physics}\ }\textbf {\bibinfo {volume} {2013}},\ \bibinfo
  {pages} {028--028} (\bibinfo {year} {2013})}\BibitemShut {NoStop}%
\bibitem [{\citenamefont {Almheiri}\ \emph
  {et~al.}(2013{\natexlab{a}})\citenamefont {Almheiri}, \citenamefont {Marolf},
  \citenamefont {Polchinski},\ and\ \citenamefont {Sully}}]{Almheiri:2012rt}%
  \BibitemOpen
  \bibfield  {author} {\bibinfo {author} {\bibfnamefont {Ahmed}\ \bibnamefont
  {Almheiri}}, \bibinfo {author} {\bibfnamefont {Donald}\ \bibnamefont
  {Marolf}}, \bibinfo {author} {\bibfnamefont {Joseph}\ \bibnamefont
  {Polchinski}}, \ and\ \bibinfo {author} {\bibfnamefont {James}\ \bibnamefont
  {Sully}},\ }\bibfield  {title} {\enquote {\bibinfo {title} {{Black Holes:
  Complementarity or Firewalls?}}}\ }\href {\doibase 10.1007/JHEP02(2013)062}
  {\bibfield  {journal} {\bibinfo  {journal} {JHEP}\ }\textbf {\bibinfo
  {volume} {02}},\ \bibinfo {pages} {062} (\bibinfo {year}
  {2013}{\natexlab{a}})},\ \Eprint {http://arxiv.org/abs/1207.3123}
  {arXiv:1207.3123 [hep-th]} \BibitemShut {NoStop}%
\bibitem [{\citenamefont {Almheiri}\ \emph
  {et~al.}(2013{\natexlab{b}})\citenamefont {Almheiri}, \citenamefont {Marolf},
  \citenamefont {Polchinski}, \citenamefont {Stanford},\ and\ \citenamefont
  {Sully}}]{Almheiri:2013hfa}%
  \BibitemOpen
  \bibfield  {author} {\bibinfo {author} {\bibfnamefont {Ahmed}\ \bibnamefont
  {Almheiri}}, \bibinfo {author} {\bibfnamefont {Donald}\ \bibnamefont
  {Marolf}}, \bibinfo {author} {\bibfnamefont {Joseph}\ \bibnamefont
  {Polchinski}}, \bibinfo {author} {\bibfnamefont {Douglas}\ \bibnamefont
  {Stanford}}, \ and\ \bibinfo {author} {\bibfnamefont {James}\ \bibnamefont
  {Sully}},\ }\bibfield  {title} {\enquote {\bibinfo {title} {{An Apologia for
  Firewalls}},}\ }\href {\doibase 10.1007/JHEP09(2013)018} {\bibfield
  {journal} {\bibinfo  {journal} {JHEP}\ }\textbf {\bibinfo {volume} {09}},\
  \bibinfo {pages} {018} (\bibinfo {year} {2013}{\natexlab{b}})},\ \Eprint
  {http://arxiv.org/abs/1304.6483} {arXiv:1304.6483 [hep-th]} \BibitemShut
  {NoStop}%
\bibitem [{\citenamefont {Lloyd}\ and\ \citenamefont
  {Preskill}(2014)}]{Lloyd:2013bza}%
  \BibitemOpen
  \bibfield  {author} {\bibinfo {author} {\bibfnamefont {Seth}\ \bibnamefont
  {Lloyd}}\ and\ \bibinfo {author} {\bibfnamefont {John}\ \bibnamefont
  {Preskill}},\ }\bibfield  {title} {\enquote {\bibinfo {title} {{Unitarity of
  black hole evaporation in final-state projection models}},}\ }\href {\doibase
  10.1007/JHEP08(2014)126} {\bibfield  {journal} {\bibinfo  {journal} {JHEP}\
  }\textbf {\bibinfo {volume} {08}},\ \bibinfo {pages} {126} (\bibinfo {year}
  {2014})},\ \Eprint {http://arxiv.org/abs/1308.4209} {arXiv:1308.4209
  [hep-th]} \BibitemShut {NoStop}%
\bibitem [{\citenamefont {Papadodimas}\ and\ \citenamefont
  {Raju}(2014)}]{Papadodimas:2013wnh}%
  \BibitemOpen
  \bibfield  {author} {\bibinfo {author} {\bibfnamefont {Kyriakos}\
  \bibnamefont {Papadodimas}}\ and\ \bibinfo {author} {\bibfnamefont {Suvrat}\
  \bibnamefont {Raju}},\ }\bibfield  {title} {\enquote {\bibinfo {title}
  {{Black Hole Interior in the Holographic Correspondence and the Information
  Paradox}},}\ }\href {\doibase 10.1103/PhysRevLett.112.051301} {\bibfield
  {journal} {\bibinfo  {journal} {Phys. Rev. Lett.}\ }\textbf {\bibinfo
  {volume} {112}},\ \bibinfo {pages} {051301} (\bibinfo {year} {2014})},\
  \Eprint {http://arxiv.org/abs/1310.6334} {arXiv:1310.6334 [hep-th]}
  \BibitemShut {NoStop}%
\bibitem [{\citenamefont {Penington}\ \emph {et~al.}(2022)\citenamefont
  {Penington}, \citenamefont {Shenker}, \citenamefont {Stanford},\ and\
  \citenamefont {Yang}}]{Penington:2019kki}%
  \BibitemOpen
  \bibfield  {author} {\bibinfo {author} {\bibfnamefont {Geoff}\ \bibnamefont
  {Penington}}, \bibinfo {author} {\bibfnamefont {Stephen~H.}\ \bibnamefont
  {Shenker}}, \bibinfo {author} {\bibfnamefont {Douglas}\ \bibnamefont
  {Stanford}}, \ and\ \bibinfo {author} {\bibfnamefont {Zhenbin}\ \bibnamefont
  {Yang}},\ }\bibfield  {title} {\enquote {\bibinfo {title} {{Replica wormholes
  and the black hole interior}},}\ }\href {\doibase 10.1007/JHEP03(2022)205}
  {\bibfield  {journal} {\bibinfo  {journal} {JHEP}\ }\textbf {\bibinfo
  {volume} {03}},\ \bibinfo {pages} {205} (\bibinfo {year} {2022})},\ \Eprint
  {http://arxiv.org/abs/1911.11977} {arXiv:1911.11977 [hep-th]} \BibitemShut
  {NoStop}%
\bibitem [{\citenamefont {Penington}(2020)}]{Penington:2019npb}%
  \BibitemOpen
  \bibfield  {author} {\bibinfo {author} {\bibfnamefont {Geoffrey}\
  \bibnamefont {Penington}},\ }\bibfield  {title} {\enquote {\bibinfo {title}
  {{Entanglement Wedge Reconstruction and the Information Paradox}},}\ }\href
  {\doibase 10.1007/JHEP09(2020)002} {\bibfield  {journal} {\bibinfo  {journal}
  {JHEP}\ }\textbf {\bibinfo {volume} {09}},\ \bibinfo {pages} {002} (\bibinfo
  {year} {2020})},\ \Eprint {http://arxiv.org/abs/1905.08255} {arXiv:1905.08255
  [hep-th]} \BibitemShut {NoStop}%
\bibitem [{\citenamefont {Almheiri}\ \emph
  {et~al.}(2020{\natexlab{a}})\citenamefont {Almheiri}, \citenamefont
  {Mahajan}, \citenamefont {Maldacena},\ and\ \citenamefont
  {Zhao}}]{Almheiri:2019hni}%
  \BibitemOpen
  \bibfield  {author} {\bibinfo {author} {\bibfnamefont {Ahmed}\ \bibnamefont
  {Almheiri}}, \bibinfo {author} {\bibfnamefont {Raghu}\ \bibnamefont
  {Mahajan}}, \bibinfo {author} {\bibfnamefont {Juan}\ \bibnamefont
  {Maldacena}}, \ and\ \bibinfo {author} {\bibfnamefont {Ying}\ \bibnamefont
  {Zhao}},\ }\bibfield  {title} {\enquote {\bibinfo {title} {{The Page curve of
  Hawking radiation from semiclassical geometry}},}\ }\href {\doibase
  10.1007/JHEP03(2020)149} {\bibfield  {journal} {\bibinfo  {journal} {JHEP}\
  }\textbf {\bibinfo {volume} {03}},\ \bibinfo {pages} {149} (\bibinfo {year}
  {2020}{\natexlab{a}})},\ \Eprint {http://arxiv.org/abs/1908.10996}
  {arXiv:1908.10996 [hep-th]} \BibitemShut {NoStop}%
\bibitem [{\citenamefont {Almheiri}\ \emph
  {et~al.}(2019{\natexlab{a}})\citenamefont {Almheiri}, \citenamefont
  {Engelhardt}, \citenamefont {Marolf},\ and\ \citenamefont
  {Maxfield}}]{Almheiri:2019psf}%
  \BibitemOpen
  \bibfield  {author} {\bibinfo {author} {\bibfnamefont {Ahmed}\ \bibnamefont
  {Almheiri}}, \bibinfo {author} {\bibfnamefont {Netta}\ \bibnamefont
  {Engelhardt}}, \bibinfo {author} {\bibfnamefont {Donald}\ \bibnamefont
  {Marolf}}, \ and\ \bibinfo {author} {\bibfnamefont {Henry}\ \bibnamefont
  {Maxfield}},\ }\bibfield  {title} {\enquote {\bibinfo {title} {{The entropy
  of bulk quantum fields and the entanglement wedge of an evaporating black
  hole}},}\ }\href {\doibase 10.1007/JHEP12(2019)063} {\bibfield  {journal}
  {\bibinfo  {journal} {JHEP}\ }\textbf {\bibinfo {volume} {12}},\ \bibinfo
  {pages} {063} (\bibinfo {year} {2019}{\natexlab{a}})},\ \Eprint
  {http://arxiv.org/abs/1905.08762} {arXiv:1905.08762 [hep-th]} \BibitemShut
  {NoStop}%
\bibitem [{\citenamefont {Almheiri}\ \emph
  {et~al.}(2020{\natexlab{b}})\citenamefont {Almheiri}, \citenamefont
  {Hartman}, \citenamefont {Maldacena}, \citenamefont {Shaghoulian},\ and\
  \citenamefont {Tajdini}}]{Almheiri:2019qdq}%
  \BibitemOpen
  \bibfield  {author} {\bibinfo {author} {\bibfnamefont {Ahmed}\ \bibnamefont
  {Almheiri}}, \bibinfo {author} {\bibfnamefont {Thomas}\ \bibnamefont
  {Hartman}}, \bibinfo {author} {\bibfnamefont {Juan}\ \bibnamefont
  {Maldacena}}, \bibinfo {author} {\bibfnamefont {Edgar}\ \bibnamefont
  {Shaghoulian}}, \ and\ \bibinfo {author} {\bibfnamefont {Amirhossein}\
  \bibnamefont {Tajdini}},\ }\bibfield  {title} {\enquote {\bibinfo {title}
  {{Replica Wormholes and the Entropy of Hawking Radiation}},}\ }\href
  {\doibase 10.1007/JHEP05(2020)013} {\bibfield  {journal} {\bibinfo  {journal}
  {JHEP}\ }\textbf {\bibinfo {volume} {05}},\ \bibinfo {pages} {013} (\bibinfo
  {year} {2020}{\natexlab{b}})},\ \Eprint {http://arxiv.org/abs/1911.12333}
  {arXiv:1911.12333 [hep-th]} \BibitemShut {NoStop}%
\bibitem [{\citenamefont {Almheiri}\ \emph
  {et~al.}(2019{\natexlab{b}})\citenamefont {Almheiri}, \citenamefont
  {Mahajan},\ and\ \citenamefont {Maldacena}}]{Almheiri:2019yqk}%
  \BibitemOpen
  \bibfield  {author} {\bibinfo {author} {\bibfnamefont {Ahmed}\ \bibnamefont
  {Almheiri}}, \bibinfo {author} {\bibfnamefont {Raghu}\ \bibnamefont
  {Mahajan}}, \ and\ \bibinfo {author} {\bibfnamefont {Juan}\ \bibnamefont
  {Maldacena}},\ }\bibfield  {title} {\enquote {\bibinfo {title} {{Islands
  outside the horizon}},}\ }\href@noop {} {\  (\bibinfo {year}
  {2019}{\natexlab{b}})},\ \Eprint {http://arxiv.org/abs/1910.11077}
  {arXiv:1910.11077 [hep-th]} \BibitemShut {NoStop}%
\bibitem [{\citenamefont {Almheiri}\ \emph {et~al.}(2021)\citenamefont
  {Almheiri}, \citenamefont {Hartman}, \citenamefont {Maldacena}, \citenamefont
  {Shaghoulian},\ and\ \citenamefont {Tajdini}}]{Almheiri:2020cfm}%
  \BibitemOpen
  \bibfield  {author} {\bibinfo {author} {\bibfnamefont {Ahmed}\ \bibnamefont
  {Almheiri}}, \bibinfo {author} {\bibfnamefont {Thomas}\ \bibnamefont
  {Hartman}}, \bibinfo {author} {\bibfnamefont {Juan}\ \bibnamefont
  {Maldacena}}, \bibinfo {author} {\bibfnamefont {Edgar}\ \bibnamefont
  {Shaghoulian}}, \ and\ \bibinfo {author} {\bibfnamefont {Amirhossein}\
  \bibnamefont {Tajdini}},\ }\bibfield  {title} {\enquote {\bibinfo {title}
  {{The entropy of Hawking radiation}},}\ }\href {\doibase
  10.1103/RevModPhys.93.035002} {\bibfield  {journal} {\bibinfo  {journal}
  {Rev. Mod. Phys.}\ }\textbf {\bibinfo {volume} {93}},\ \bibinfo {pages}
  {035002} (\bibinfo {year} {2021})},\ \Eprint
  {http://arxiv.org/abs/2006.06872} {arXiv:2006.06872 [hep-th]} \BibitemShut
  {NoStop}%
\bibitem [{\citenamefont {Engelhardt}\ and\ \citenamefont
  {Wall}(2015)}]{Engelhardt:2014gca}%
  \BibitemOpen
  \bibfield  {author} {\bibinfo {author} {\bibfnamefont {Netta}\ \bibnamefont
  {Engelhardt}}\ and\ \bibinfo {author} {\bibfnamefont {Aron~C.}\ \bibnamefont
  {Wall}},\ }\bibfield  {title} {\enquote {\bibinfo {title} {{Quantum Extremal
  Surfaces: Holographic Entanglement Entropy beyond the Classical Regime}},}\
  }\href {\doibase 10.1007/JHEP01(2015)073} {\bibfield  {journal} {\bibinfo
  {journal} {JHEP}\ }\textbf {\bibinfo {volume} {01}},\ \bibinfo {pages} {073}
  (\bibinfo {year} {2015})},\ \Eprint {http://arxiv.org/abs/1408.3203}
  {arXiv:1408.3203 [hep-th]} \BibitemShut {NoStop}%
\bibitem [{\citenamefont {Engelhardt}\ and\ \citenamefont
  {Fischetti}(2019)}]{Engelhardt:2019hmr}%
  \BibitemOpen
  \bibfield  {author} {\bibinfo {author} {\bibfnamefont {Netta}\ \bibnamefont
  {Engelhardt}}\ and\ \bibinfo {author} {\bibfnamefont {Sebastian}\
  \bibnamefont {Fischetti}},\ }\bibfield  {title} {\enquote {\bibinfo {title}
  {{Surface Theory: the Classical, the Quantum, and the Holographic}},}\ }\href
  {\doibase 10.1088/1361-6382/ab3bda} {\bibfield  {journal} {\bibinfo
  {journal} {Class. Quant. Grav.}\ }\textbf {\bibinfo {volume} {36}},\ \bibinfo
  {pages} {205002} (\bibinfo {year} {2019})},\ \Eprint
  {http://arxiv.org/abs/1904.08423} {arXiv:1904.08423 [hep-th]} \BibitemShut
  {NoStop}%
\bibitem [{\citenamefont {Akers}\ \emph {et~al.}(2020)\citenamefont {Akers},
  \citenamefont {Engelhardt}, \citenamefont {Penington},\ and\ \citenamefont
  {Usatyuk}}]{Akers:2019lzs}%
  \BibitemOpen
  \bibfield  {author} {\bibinfo {author} {\bibfnamefont {Chris}\ \bibnamefont
  {Akers}}, \bibinfo {author} {\bibfnamefont {Netta}\ \bibnamefont
  {Engelhardt}}, \bibinfo {author} {\bibfnamefont {Geoff}\ \bibnamefont
  {Penington}}, \ and\ \bibinfo {author} {\bibfnamefont {Mykhaylo}\
  \bibnamefont {Usatyuk}},\ }\bibfield  {title} {\enquote {\bibinfo {title}
  {{Quantum Maximin Surfaces}},}\ }\href {\doibase 10.1007/JHEP08(2020)140}
  {\bibfield  {journal} {\bibinfo  {journal} {JHEP}\ }\textbf {\bibinfo
  {volume} {08}},\ \bibinfo {pages} {140} (\bibinfo {year} {2020})},\ \Eprint
  {http://arxiv.org/abs/1912.02799} {arXiv:1912.02799 [hep-th]} \BibitemShut
  {NoStop}%
\bibitem [{\citenamefont {Wall}(2014)}]{Wall:2012uf}%
  \BibitemOpen
  \bibfield  {author} {\bibinfo {author} {\bibfnamefont {Aron~C.}\ \bibnamefont
  {Wall}},\ }\bibfield  {title} {\enquote {\bibinfo {title} {{Maximin Surfaces,
  and the Strong Subadditivity of the Covariant Holographic Entanglement
  Entropy}},}\ }\href {\doibase 10.1088/0264-9381/31/22/225007} {\bibfield
  {journal} {\bibinfo  {journal} {Class. Quant. Grav.}\ }\textbf {\bibinfo
  {volume} {31}},\ \bibinfo {pages} {225007} (\bibinfo {year} {2014})},\
  \Eprint {http://arxiv.org/abs/1211.3494} {arXiv:1211.3494 [hep-th]}
  \BibitemShut {NoStop}%
\bibitem [{\citenamefont {Kundu}(2022)}]{Kundu:2021nwp}%
  \BibitemOpen
  \bibfield  {author} {\bibinfo {author} {\bibfnamefont {Arnab}\ \bibnamefont
  {Kundu}},\ }\bibfield  {title} {\enquote {\bibinfo {title} {{Wormholes and
  holography: an introduction}},}\ }\href {\doibase
  10.1140/epjc/s10052-022-10376-z} {\bibfield  {journal} {\bibinfo  {journal}
  {Eur. Phys. J. C}\ }\textbf {\bibinfo {volume} {82}},\ \bibinfo {pages} {447}
  (\bibinfo {year} {2022})},\ \Eprint {http://arxiv.org/abs/2110.14958}
  {arXiv:2110.14958 [hep-th]} \BibitemShut {NoStop}%
\bibitem [{\citenamefont {Goto}\ \emph {et~al.}(2021)\citenamefont {Goto},
  \citenamefont {Hartman},\ and\ \citenamefont {Tajdini}}]{Goto:2020wnk}%
  \BibitemOpen
  \bibfield  {author} {\bibinfo {author} {\bibfnamefont {Kanato}\ \bibnamefont
  {Goto}}, \bibinfo {author} {\bibfnamefont {Thomas}\ \bibnamefont {Hartman}},
  \ and\ \bibinfo {author} {\bibfnamefont {Amirhossein}\ \bibnamefont
  {Tajdini}},\ }\bibfield  {title} {\enquote {\bibinfo {title} {{Replica
  wormholes for an evaporating 2D black hole}},}\ }\href {\doibase
  10.1007/JHEP04(2021)289} {\bibfield  {journal} {\bibinfo  {journal} {JHEP}\
  }\textbf {\bibinfo {volume} {04}},\ \bibinfo {pages} {289} (\bibinfo {year}
  {2021})},\ \Eprint {http://arxiv.org/abs/2011.09043} {arXiv:2011.09043
  [hep-th]} \BibitemShut {NoStop}%
\bibitem [{\citenamefont {Colin-Ellerin}\ \emph {et~al.}(2021)\citenamefont
  {Colin-Ellerin}, \citenamefont {Dong}, \citenamefont {Marolf}, \citenamefont
  {Rangamani},\ and\ \citenamefont {Wang}}]{Colin-Ellerin:2020mva}%
  \BibitemOpen
  \bibfield  {author} {\bibinfo {author} {\bibfnamefont {Sean}\ \bibnamefont
  {Colin-Ellerin}}, \bibinfo {author} {\bibfnamefont {Xi}~\bibnamefont {Dong}},
  \bibinfo {author} {\bibfnamefont {Donald}\ \bibnamefont {Marolf}}, \bibinfo
  {author} {\bibfnamefont {Mukund}\ \bibnamefont {Rangamani}}, \ and\ \bibinfo
  {author} {\bibfnamefont {Zhencheng}\ \bibnamefont {Wang}},\ }\bibfield
  {title} {\enquote {\bibinfo {title} {{Real-time gravitational replicas:
  Formalism and a variational principle}},}\ }\href {\doibase
  10.1007/JHEP05(2021)117} {\bibfield  {journal} {\bibinfo  {journal} {JHEP}\
  }\textbf {\bibinfo {volume} {05}},\ \bibinfo {pages} {117} (\bibinfo {year}
  {2021})},\ \Eprint {http://arxiv.org/abs/2012.00828} {arXiv:2012.00828
  [hep-th]} \BibitemShut {NoStop}%
\bibitem [{\citenamefont {Chen}\ \emph {et~al.}(2020)\citenamefont {Chen},
  \citenamefont {Fisher}, \citenamefont {Hernandez}, \citenamefont {Myers},\
  and\ \citenamefont {Ruan}}]{Chen:2019uhq}%
  \BibitemOpen
  \bibfield  {author} {\bibinfo {author} {\bibfnamefont {Hong~Zhe}\
  \bibnamefont {Chen}}, \bibinfo {author} {\bibfnamefont {Zachary}\
  \bibnamefont {Fisher}}, \bibinfo {author} {\bibfnamefont {Juan}\ \bibnamefont
  {Hernandez}}, \bibinfo {author} {\bibfnamefont {Robert~C.}\ \bibnamefont
  {Myers}}, \ and\ \bibinfo {author} {\bibfnamefont {Shan-Ming}\ \bibnamefont
  {Ruan}},\ }\bibfield  {title} {\enquote {\bibinfo {title} {{Information Flow
  in Black Hole Evaporation}},}\ }\href {\doibase 10.1007/JHEP03(2020)152}
  {\bibfield  {journal} {\bibinfo  {journal} {JHEP}\ }\textbf {\bibinfo
  {volume} {03}},\ \bibinfo {pages} {152} (\bibinfo {year} {2020})},\ \Eprint
  {http://arxiv.org/abs/1911.03402} {arXiv:1911.03402 [hep-th]} \BibitemShut
  {NoStop}%
\bibitem [{\citenamefont {Hashimoto}\ \emph {et~al.}(2020)\citenamefont
  {Hashimoto}, \citenamefont {Iizuka},\ and\ \citenamefont
  {Matsuo}}]{Hashimoto:2020cas}%
  \BibitemOpen
  \bibfield  {author} {\bibinfo {author} {\bibfnamefont {Koji}\ \bibnamefont
  {Hashimoto}}, \bibinfo {author} {\bibfnamefont {Norihiro}\ \bibnamefont
  {Iizuka}}, \ and\ \bibinfo {author} {\bibfnamefont {Yoshinori}\ \bibnamefont
  {Matsuo}},\ }\bibfield  {title} {\enquote {\bibinfo {title} {{Islands in
  Schwarzschild black holes}},}\ }\href {\doibase 10.1007/JHEP06(2020)085}
  {\bibfield  {journal} {\bibinfo  {journal} {JHEP}\ }\textbf {\bibinfo
  {volume} {06}},\ \bibinfo {pages} {085} (\bibinfo {year} {2020})},\ \Eprint
  {http://arxiv.org/abs/2004.05863} {arXiv:2004.05863 [hep-th]} \BibitemShut
  {NoStop}%
\bibitem [{\citenamefont {Hartman}\ \emph {et~al.}(2020)\citenamefont
  {Hartman}, \citenamefont {Shaghoulian},\ and\ \citenamefont
  {Strominger}}]{Hartman:2020swn}%
  \BibitemOpen
  \bibfield  {author} {\bibinfo {author} {\bibfnamefont {Thomas}\ \bibnamefont
  {Hartman}}, \bibinfo {author} {\bibfnamefont {Edgar}\ \bibnamefont
  {Shaghoulian}}, \ and\ \bibinfo {author} {\bibfnamefont {Andrew}\
  \bibnamefont {Strominger}},\ }\bibfield  {title} {\enquote {\bibinfo {title}
  {{Islands in Asymptotically Flat 2D Gravity}},}\ }\href {\doibase
  10.1007/JHEP07(2020)022} {\bibfield  {journal} {\bibinfo  {journal} {JHEP}\
  }\textbf {\bibinfo {volume} {07}},\ \bibinfo {pages} {022} (\bibinfo {year}
  {2020})},\ \Eprint {http://arxiv.org/abs/2004.13857} {arXiv:2004.13857
  [hep-th]} \BibitemShut {NoStop}%
\bibitem [{\citenamefont {Anegawa}\ and\ \citenamefont
  {Iizuka}(2020)}]{Anegawa:2020ezn}%
  \BibitemOpen
  \bibfield  {author} {\bibinfo {author} {\bibfnamefont {Takanori}\
  \bibnamefont {Anegawa}}\ and\ \bibinfo {author} {\bibfnamefont {Norihiro}\
  \bibnamefont {Iizuka}},\ }\bibfield  {title} {\enquote {\bibinfo {title}
  {{Notes on islands in asymptotically flat 2d dilaton black holes}},}\ }\href
  {\doibase 10.1007/JHEP07(2020)036} {\bibfield  {journal} {\bibinfo  {journal}
  {JHEP}\ }\textbf {\bibinfo {volume} {07}},\ \bibinfo {pages} {036} (\bibinfo
  {year} {2020})},\ \Eprint {http://arxiv.org/abs/2004.01601} {arXiv:2004.01601
  [hep-th]} \BibitemShut {NoStop}%
\bibitem [{\citenamefont {Dong}\ \emph {et~al.}(2020)\citenamefont {Dong},
  \citenamefont {Qi}, \citenamefont {Shangnan},\ and\ \citenamefont
  {Yang}}]{Dong:2020uxp}%
  \BibitemOpen
  \bibfield  {author} {\bibinfo {author} {\bibfnamefont {Xi}~\bibnamefont
  {Dong}}, \bibinfo {author} {\bibfnamefont {Xiao-Liang}\ \bibnamefont {Qi}},
  \bibinfo {author} {\bibfnamefont {Zhou}\ \bibnamefont {Shangnan}}, \ and\
  \bibinfo {author} {\bibfnamefont {Zhenbin}\ \bibnamefont {Yang}},\ }\bibfield
   {title} {\enquote {\bibinfo {title} {{Effective entropy of quantum fields
  coupled with gravity}},}\ }\href {\doibase 10.1007/JHEP10(2020)052}
  {\bibfield  {journal} {\bibinfo  {journal} {JHEP}\ }\textbf {\bibinfo
  {volume} {10}},\ \bibinfo {pages} {052} (\bibinfo {year} {2020})},\ \Eprint
  {http://arxiv.org/abs/2007.02987} {arXiv:2007.02987 [hep-th]} \BibitemShut
  {NoStop}%
\bibitem [{\citenamefont {Balasubramanian}\ \emph {et~al.}(2021)\citenamefont
  {Balasubramanian}, \citenamefont {Kar},\ and\ \citenamefont
  {Ugajin}}]{Balasubramanian:2020xqf}%
  \BibitemOpen
  \bibfield  {author} {\bibinfo {author} {\bibfnamefont {Vijay}\ \bibnamefont
  {Balasubramanian}}, \bibinfo {author} {\bibfnamefont {Arjun}\ \bibnamefont
  {Kar}}, \ and\ \bibinfo {author} {\bibfnamefont {Tomonori}\ \bibnamefont
  {Ugajin}},\ }\bibfield  {title} {\enquote {\bibinfo {title} {{Islands in de
  Sitter space}},}\ }\href {\doibase 10.1007/JHEP02(2021)072} {\bibfield
  {journal} {\bibinfo  {journal} {JHEP}\ }\textbf {\bibinfo {volume} {02}},\
  \bibinfo {pages} {072} (\bibinfo {year} {2021})},\ \Eprint
  {http://arxiv.org/abs/2008.05275} {arXiv:2008.05275 [hep-th]} \BibitemShut
  {NoStop}%
\bibitem [{\citenamefont {Alishahiha}\ \emph
  {et~al.}(2021{\natexlab{a}})\citenamefont {Alishahiha}, \citenamefont
  {Faraji~Astaneh},\ and\ \citenamefont {Naseh}}]{Alishahiha:2020qza}%
  \BibitemOpen
  \bibfield  {author} {\bibinfo {author} {\bibfnamefont {Mohsen}\ \bibnamefont
  {Alishahiha}}, \bibinfo {author} {\bibfnamefont {Amin}\ \bibnamefont
  {Faraji~Astaneh}}, \ and\ \bibinfo {author} {\bibfnamefont {Ali}\
  \bibnamefont {Naseh}},\ }\bibfield  {title} {\enquote {\bibinfo {title}
  {{Island in the presence of higher derivative terms}},}\ }\href {\doibase
  10.1007/JHEP02(2021)035} {\bibfield  {journal} {\bibinfo  {journal} {JHEP}\
  }\textbf {\bibinfo {volume} {02}},\ \bibinfo {pages} {035} (\bibinfo {year}
  {2021}{\natexlab{a}})},\ \Eprint {http://arxiv.org/abs/2005.08715}
  {arXiv:2005.08715 [hep-th]} \BibitemShut {NoStop}%
\bibitem [{\citenamefont {Azarnia}\ \emph {et~al.}(2021)\citenamefont
  {Azarnia}, \citenamefont {Fareghbal}, \citenamefont {Naseh},\ and\
  \citenamefont {Zolfi}}]{Azarnia:2021uch}%
  \BibitemOpen
  \bibfield  {author} {\bibinfo {author} {\bibfnamefont {Sanam}\ \bibnamefont
  {Azarnia}}, \bibinfo {author} {\bibfnamefont {Reza}\ \bibnamefont
  {Fareghbal}}, \bibinfo {author} {\bibfnamefont {Ali}\ \bibnamefont {Naseh}},
  \ and\ \bibinfo {author} {\bibfnamefont {Hamed}\ \bibnamefont {Zolfi}},\
  }\bibfield  {title} {\enquote {\bibinfo {title} {{Islands in flat-space
  cosmology}},}\ }\href {\doibase 10.1103/PhysRevD.104.126017} {\bibfield
  {journal} {\bibinfo  {journal} {Phys. Rev. D}\ }\textbf {\bibinfo {volume}
  {104}},\ \bibinfo {pages} {126017} (\bibinfo {year} {2021})},\ \Eprint
  {http://arxiv.org/abs/2109.04795} {arXiv:2109.04795 [hep-th]} \BibitemShut
  {NoStop}%
\bibitem [{\citenamefont {Aref'eva}\ and\ \citenamefont
  {Volovich}(2021)}]{Arefeva:2021kfx}%
  \BibitemOpen
  \bibfield  {author} {\bibinfo {author} {\bibfnamefont {Irina}\ \bibnamefont
  {Aref'eva}}\ and\ \bibinfo {author} {\bibfnamefont {Igor}\ \bibnamefont
  {Volovich}},\ }\bibfield  {title} {\enquote {\bibinfo {title} {{A Note on
  Islands in Schwarzschild Black Holes}},}\ }\href@noop {} {\  (\bibinfo {year}
  {2021})},\ \Eprint {http://arxiv.org/abs/2110.04233} {arXiv:2110.04233
  [hep-th]} \BibitemShut {NoStop}%
\bibitem [{\citenamefont {He}\ \emph {et~al.}(2022)\citenamefont {He},
  \citenamefont {Sun}, \citenamefont {Zhao},\ and\ \citenamefont
  {Zhang}}]{He:2021mst}%
  \BibitemOpen
  \bibfield  {author} {\bibinfo {author} {\bibfnamefont {Song}\ \bibnamefont
  {He}}, \bibinfo {author} {\bibfnamefont {Yuan}\ \bibnamefont {Sun}}, \bibinfo
  {author} {\bibfnamefont {Long}\ \bibnamefont {Zhao}}, \ and\ \bibinfo
  {author} {\bibfnamefont {Yu-Xuan}\ \bibnamefont {Zhang}},\ }\bibfield
  {title} {\enquote {\bibinfo {title} {{The universality of islands outside the
  horizon}},}\ }\href {\doibase 10.1007/JHEP05(2022)047} {\bibfield  {journal}
  {\bibinfo  {journal} {JHEP}\ }\textbf {\bibinfo {volume} {05}},\ \bibinfo
  {pages} {047} (\bibinfo {year} {2022})},\ \Eprint
  {http://arxiv.org/abs/2110.07598} {arXiv:2110.07598 [hep-th]} \BibitemShut
  {NoStop}%
\bibitem [{\citenamefont {Omidi}(2022)}]{Omidi:2021opl}%
  \BibitemOpen
  \bibfield  {author} {\bibinfo {author} {\bibfnamefont {Farzad}\ \bibnamefont
  {Omidi}},\ }\bibfield  {title} {\enquote {\bibinfo {title} {{Entropy of
  Hawking radiation for two-sided hyperscaling violating black branes}},}\
  }\href {\doibase 10.1007/JHEP04(2022)022} {\bibfield  {journal} {\bibinfo
  {journal} {JHEP}\ }\textbf {\bibinfo {volume} {04}},\ \bibinfo {pages} {022}
  (\bibinfo {year} {2022})},\ \Eprint {http://arxiv.org/abs/2112.05890}
  {arXiv:2112.05890 [hep-th]} \BibitemShut {NoStop}%
\bibitem [{\citenamefont {Yu}\ \emph {et~al.}(2022)\citenamefont {Yu},
  \citenamefont {Lu}, \citenamefont {Ge},\ and\ \citenamefont
  {Sin}}]{Yu:2021rfg}%
  \BibitemOpen
  \bibfield  {author} {\bibinfo {author} {\bibfnamefont {Ming-Hui}\
  \bibnamefont {Yu}}, \bibinfo {author} {\bibfnamefont {Cheng-Yuan}\
  \bibnamefont {Lu}}, \bibinfo {author} {\bibfnamefont {Xian-Hui}\ \bibnamefont
  {Ge}}, \ and\ \bibinfo {author} {\bibfnamefont {Sang-Jin}\ \bibnamefont
  {Sin}},\ }\bibfield  {title} {\enquote {\bibinfo {title} {{Island, Page
  curve, and superradiance of rotating BTZ black holes}},}\ }\href {\doibase
  10.1103/PhysRevD.105.066009} {\bibfield  {journal} {\bibinfo  {journal}
  {Phys. Rev. D}\ }\textbf {\bibinfo {volume} {105}},\ \bibinfo {pages}
  {066009} (\bibinfo {year} {2022})},\ \Eprint
  {http://arxiv.org/abs/2112.14361} {arXiv:2112.14361 [hep-th]} \BibitemShut
  {NoStop}%
\bibitem [{\citenamefont {Yadav}(2022)}]{Yadav:2022fmo}%
  \BibitemOpen
  \bibfield  {author} {\bibinfo {author} {\bibfnamefont {Gopal}\ \bibnamefont
  {Yadav}},\ }\bibfield  {title} {\enquote {\bibinfo {title} {{Page Curves of
  Reissner-Nordstr\"om Black Hole in HD Gravity}},}\ }\href@noop {} {\
  (\bibinfo {year} {2022})},\ \Eprint {http://arxiv.org/abs/2204.11882}
  {arXiv:2204.11882 [hep-th]} \BibitemShut {NoStop}%
\bibitem [{\citenamefont {Du}\ \emph {et~al.}(2022)\citenamefont {Du},
  \citenamefont {Gan}, \citenamefont {Shu},\ and\ \citenamefont
  {Sun}}]{Du:2022vvg}%
  \BibitemOpen
  \bibfield  {author} {\bibinfo {author} {\bibfnamefont {Dong-Hui}\
  \bibnamefont {Du}}, \bibinfo {author} {\bibfnamefont {Wen-Cong}\ \bibnamefont
  {Gan}}, \bibinfo {author} {\bibfnamefont {Fu-Wen}\ \bibnamefont {Shu}}, \
  and\ \bibinfo {author} {\bibfnamefont {Jia-Rui}\ \bibnamefont {Sun}},\
  }\bibfield  {title} {\enquote {\bibinfo {title} {{Unitary Constraints on
  Semiclassical Schwarzschild Black Holes in the Presence of Island}},}\
  }\href@noop {} {\  (\bibinfo {year} {2022})},\ \Eprint
  {http://arxiv.org/abs/2206.10339} {arXiv:2206.10339 [hep-th]} \BibitemShut
  {NoStop}%
\bibitem [{\citenamefont {Geng}\ and\ \citenamefont
  {Karch}(2020)}]{Geng:2020qvw}%
  \BibitemOpen
  \bibfield  {author} {\bibinfo {author} {\bibfnamefont {Hao}\ \bibnamefont
  {Geng}}\ and\ \bibinfo {author} {\bibfnamefont {Andreas}\ \bibnamefont
  {Karch}},\ }\bibfield  {title} {\enquote {\bibinfo {title} {{Massive
  islands}},}\ }\href {\doibase 10.1007/JHEP09(2020)121} {\bibfield  {journal}
  {\bibinfo  {journal} {JHEP}\ }\textbf {\bibinfo {volume} {09}},\ \bibinfo
  {pages} {121} (\bibinfo {year} {2020})},\ \Eprint
  {http://arxiv.org/abs/2006.02438} {arXiv:2006.02438 [hep-th]} \BibitemShut
  {NoStop}%
\bibitem [{\citenamefont {Geng}\ \emph
  {et~al.}(2022{\natexlab{a}})\citenamefont {Geng}, \citenamefont {Karch},
  \citenamefont {Perez-Pardavila}, \citenamefont {Raju}, \citenamefont
  {Randall}, \citenamefont {Riojas},\ and\ \citenamefont
  {Shashi}}]{Geng:2021hlu}%
  \BibitemOpen
  \bibfield  {author} {\bibinfo {author} {\bibfnamefont {Hao}\ \bibnamefont
  {Geng}}, \bibinfo {author} {\bibfnamefont {Andreas}\ \bibnamefont {Karch}},
  \bibinfo {author} {\bibfnamefont {Carlos}\ \bibnamefont {Perez-Pardavila}},
  \bibinfo {author} {\bibfnamefont {Suvrat}\ \bibnamefont {Raju}}, \bibinfo
  {author} {\bibfnamefont {Lisa}\ \bibnamefont {Randall}}, \bibinfo {author}
  {\bibfnamefont {Marcos}\ \bibnamefont {Riojas}}, \ and\ \bibinfo {author}
  {\bibfnamefont {Sanjit}\ \bibnamefont {Shashi}},\ }\bibfield  {title}
  {\enquote {\bibinfo {title} {{Inconsistency of islands in theories with
  long-range gravity}},}\ }\href {\doibase 10.1007/JHEP01(2022)182} {\bibfield
  {journal} {\bibinfo  {journal} {JHEP}\ }\textbf {\bibinfo {volume} {01}},\
  \bibinfo {pages} {182} (\bibinfo {year} {2022}{\natexlab{a}})},\ \Eprint
  {http://arxiv.org/abs/2107.03390} {arXiv:2107.03390 [hep-th]} \BibitemShut
  {NoStop}%
\bibitem [{\citenamefont {Hollowood}\ and\ \citenamefont
  {Kumar}(2020)}]{Hollowood:2020cou}%
  \BibitemOpen
  \bibfield  {author} {\bibinfo {author} {\bibfnamefont {Timothy~J.}\
  \bibnamefont {Hollowood}}\ and\ \bibinfo {author} {\bibfnamefont {S.~Prem}\
  \bibnamefont {Kumar}},\ }\bibfield  {title} {\enquote {\bibinfo {title}
  {{Islands and Page Curves for Evaporating Black Holes in JT Gravity}},}\
  }\href {\doibase 10.1007/JHEP08(2020)094} {\bibfield  {journal} {\bibinfo
  {journal} {JHEP}\ }\textbf {\bibinfo {volume} {08}},\ \bibinfo {pages} {094}
  (\bibinfo {year} {2020})},\ \Eprint {http://arxiv.org/abs/2004.14944}
  {arXiv:2004.14944 [hep-th]} \BibitemShut {NoStop}%
\bibitem [{\citenamefont {Wang}\ \emph
  {et~al.}(2021{\natexlab{a}})\citenamefont {Wang}, \citenamefont {Li},\ and\
  \citenamefont {Wang}}]{Wang:2021mqq}%
  \BibitemOpen
  \bibfield  {author} {\bibinfo {author} {\bibfnamefont {Xuanhua}\ \bibnamefont
  {Wang}}, \bibinfo {author} {\bibfnamefont {Ran}\ \bibnamefont {Li}}, \ and\
  \bibinfo {author} {\bibfnamefont {Jin}\ \bibnamefont {Wang}},\ }\bibfield
  {title} {\enquote {\bibinfo {title} {{Page curves for a family of exactly
  solvable evaporating black holes}},}\ }\href {\doibase
  10.1103/PhysRevD.103.126026} {\bibfield  {journal} {\bibinfo  {journal}
  {Phys. Rev. D}\ }\textbf {\bibinfo {volume} {103}},\ \bibinfo {pages}
  {126026} (\bibinfo {year} {2021}{\natexlab{a}})},\ \Eprint
  {http://arxiv.org/abs/2104.00224} {arXiv:2104.00224 [hep-th]} \BibitemShut
  {NoStop}%
\bibitem [{\citenamefont {Li}\ \emph {et~al.}(2021)\citenamefont {Li},
  \citenamefont {Wang},\ and\ \citenamefont {Wang}}]{Li:2021lfo}%
  \BibitemOpen
  \bibfield  {author} {\bibinfo {author} {\bibfnamefont {Ran}\ \bibnamefont
  {Li}}, \bibinfo {author} {\bibfnamefont {Xuanhua}\ \bibnamefont {Wang}}, \
  and\ \bibinfo {author} {\bibfnamefont {Jin}\ \bibnamefont {Wang}},\
  }\bibfield  {title} {\enquote {\bibinfo {title} {{Island may not save the
  information paradox of Liouville black holes}},}\ }\href {\doibase
  10.1103/PhysRevD.104.106015} {\bibfield  {journal} {\bibinfo  {journal}
  {Phys. Rev. D}\ }\textbf {\bibinfo {volume} {104}},\ \bibinfo {pages}
  {106015} (\bibinfo {year} {2021})},\ \Eprint
  {http://arxiv.org/abs/2105.03271} {arXiv:2105.03271 [hep-th]} \BibitemShut
  {NoStop}%
\bibitem [{\citenamefont {Pedraza}\ \emph {et~al.}(2021)\citenamefont
  {Pedraza}, \citenamefont {Svesko}, \citenamefont {Sybesma},\ and\
  \citenamefont {Visser}}]{Pedraza:2021cvx}%
  \BibitemOpen
  \bibfield  {author} {\bibinfo {author} {\bibfnamefont {Juan~F.}\ \bibnamefont
  {Pedraza}}, \bibinfo {author} {\bibfnamefont {Andrew}\ \bibnamefont
  {Svesko}}, \bibinfo {author} {\bibfnamefont {Watse}\ \bibnamefont {Sybesma}},
  \ and\ \bibinfo {author} {\bibfnamefont {Manus~R.}\ \bibnamefont {Visser}},\
  }\bibfield  {title} {\enquote {\bibinfo {title} {{Semi-classical
  thermodynamics of quantum extremal surfaces in Jackiw-Teitelboim gravity}},}\
  }\href {\doibase 10.1007/JHEP12(2021)134} {\bibfield  {journal} {\bibinfo
  {journal} {JHEP}\ }\textbf {\bibinfo {volume} {12}},\ \bibinfo {pages} {134}
  (\bibinfo {year} {2021})},\ \Eprint {http://arxiv.org/abs/2107.10358}
  {arXiv:2107.10358 [hep-th]} \BibitemShut {NoStop}%
\bibitem [{\citenamefont {Almheiri}\ \emph
  {et~al.}(2020{\natexlab{c}})\citenamefont {Almheiri}, \citenamefont
  {Mahajan},\ and\ \citenamefont {Santos}}]{Almheiri:2019psy}%
  \BibitemOpen
  \bibfield  {author} {\bibinfo {author} {\bibfnamefont {Ahmed}\ \bibnamefont
  {Almheiri}}, \bibinfo {author} {\bibfnamefont {Raghu}\ \bibnamefont
  {Mahajan}}, \ and\ \bibinfo {author} {\bibfnamefont {Jorge~E.}\ \bibnamefont
  {Santos}},\ }\bibfield  {title} {\enquote {\bibinfo {title} {{Entanglement
  islands in higher dimensions}},}\ }\href {\doibase
  10.21468/SciPostPhys.9.1.001} {\bibfield  {journal} {\bibinfo  {journal}
  {SciPost Phys.}\ }\textbf {\bibinfo {volume} {9}},\ \bibinfo {pages} {001}
  (\bibinfo {year} {2020}{\natexlab{c}})},\ \Eprint
  {http://arxiv.org/abs/1911.09666} {arXiv:1911.09666 [hep-th]} \BibitemShut
  {NoStop}%
\bibitem [{\citenamefont {Matsuo}(2021)}]{Matsuo:2020ypv}%
  \BibitemOpen
  \bibfield  {author} {\bibinfo {author} {\bibfnamefont {Y.}~\bibnamefont
  {Matsuo}},\ }\bibfield  {title} {\enquote {\bibinfo {title} {{Islands and
  stretched horizon}},}\ }\href {\doibase 10.1007/JHEP07(2021)051} {\bibfield
  {journal} {\bibinfo  {journal} {JHEP}\ }\textbf {\bibinfo {volume} {07}},\
  \bibinfo {pages} {051} (\bibinfo {year} {2021})},\ \Eprint
  {http://arxiv.org/abs/2011.08814} {arXiv:2011.08814 [hep-th]} \BibitemShut
  {NoStop}%
\bibitem [{\citenamefont {Krishnan}(2021)}]{Krishnan:2020fer}%
  \BibitemOpen
  \bibfield  {author} {\bibinfo {author} {\bibfnamefont {Chethan}\ \bibnamefont
  {Krishnan}},\ }\bibfield  {title} {\enquote {\bibinfo {title} {{Critical
  Islands}},}\ }\href {\doibase 10.1007/JHEP01(2021)179} {\bibfield  {journal}
  {\bibinfo  {journal} {JHEP}\ }\textbf {\bibinfo {volume} {01}},\ \bibinfo
  {pages} {179} (\bibinfo {year} {2021})},\ \Eprint
  {http://arxiv.org/abs/2007.06551} {arXiv:2007.06551 [hep-th]} \BibitemShut
  {NoStop}%
\bibitem [{\citenamefont {Krishnan}\ \emph {et~al.}(2020)\citenamefont
  {Krishnan}, \citenamefont {Patil},\ and\ \citenamefont
  {Pereira}}]{Krishnan:2020oun}%
  \BibitemOpen
  \bibfield  {author} {\bibinfo {author} {\bibfnamefont {Chethan}\ \bibnamefont
  {Krishnan}}, \bibinfo {author} {\bibfnamefont {Vaishnavi}\ \bibnamefont
  {Patil}}, \ and\ \bibinfo {author} {\bibfnamefont {Jude}\ \bibnamefont
  {Pereira}},\ }\bibfield  {title} {\enquote {\bibinfo {title} {{Page Curve and
  the Information Paradox in Flat Space}},}\ }\href@noop {} {\  (\bibinfo
  {year} {2020})},\ \Eprint {http://arxiv.org/abs/2005.02993} {arXiv:2005.02993
  [hep-th]} \BibitemShut {NoStop}%
\bibitem [{\citenamefont {Geng}\ \emph
  {et~al.}(2021{\natexlab{a}})\citenamefont {Geng}, \citenamefont {L\"ust},
  \citenamefont {Mishra},\ and\ \citenamefont {Wakeham}}]{Geng:2021iyq}%
  \BibitemOpen
  \bibfield  {author} {\bibinfo {author} {\bibfnamefont {Hao}\ \bibnamefont
  {Geng}}, \bibinfo {author} {\bibfnamefont {Severin}\ \bibnamefont {L\"ust}},
  \bibinfo {author} {\bibfnamefont {Rashmish~K.}\ \bibnamefont {Mishra}}, \
  and\ \bibinfo {author} {\bibfnamefont {David}\ \bibnamefont {Wakeham}},\
  }\bibfield  {title} {\enquote {\bibinfo {title} {{Holographic BCFTs and
  Communicating Black Holes}},}\ }\href {\doibase 10.1007/JHEP08(2021)003}
  {\bibfield  {journal} {\bibinfo  {journal} {JHEP}\ }\textbf {\bibinfo
  {volume} {08}},\ \bibinfo {pages} {003} (\bibinfo {year}
  {2021}{\natexlab{a}})},\ \Eprint {http://arxiv.org/abs/2104.07039}
  {arXiv:2104.07039 [hep-th]} \BibitemShut {NoStop}%
\bibitem [{\citenamefont {Geng}\ \emph
  {et~al.}(2022{\natexlab{b}})\citenamefont {Geng}, \citenamefont {Karch},
  \citenamefont {Perez-Pardavila}, \citenamefont {Raju}, \citenamefont
  {Randall}, \citenamefont {Riojas},\ and\ \citenamefont
  {Shashi}}]{Geng:2021mic}%
  \BibitemOpen
  \bibfield  {author} {\bibinfo {author} {\bibfnamefont {Hao}\ \bibnamefont
  {Geng}}, \bibinfo {author} {\bibfnamefont {Andreas}\ \bibnamefont {Karch}},
  \bibinfo {author} {\bibfnamefont {Carlos}\ \bibnamefont {Perez-Pardavila}},
  \bibinfo {author} {\bibfnamefont {Suvrat}\ \bibnamefont {Raju}}, \bibinfo
  {author} {\bibfnamefont {Lisa}\ \bibnamefont {Randall}}, \bibinfo {author}
  {\bibfnamefont {Marcos}\ \bibnamefont {Riojas}}, \ and\ \bibinfo {author}
  {\bibfnamefont {Sanjit}\ \bibnamefont {Shashi}},\ }\bibfield  {title}
  {\enquote {\bibinfo {title} {{Entanglement phase structure of a holographic
  BCFT in a black hole background}},}\ }\href {\doibase
  10.1007/JHEP05(2022)153} {\bibfield  {journal} {\bibinfo  {journal} {JHEP}\
  }\textbf {\bibinfo {volume} {05}},\ \bibinfo {pages} {153} (\bibinfo {year}
  {2022}{\natexlab{b}})},\ \Eprint {http://arxiv.org/abs/2112.09132}
  {arXiv:2112.09132 [hep-th]} \BibitemShut {NoStop}%
\bibitem [{\citenamefont {Goswami}\ and\ \citenamefont
  {Narayan}(2022)}]{Goswami:2022ylc}%
  \BibitemOpen
  \bibfield  {author} {\bibinfo {author} {\bibfnamefont {Kaberi}\ \bibnamefont
  {Goswami}}\ and\ \bibinfo {author} {\bibfnamefont {K.}~\bibnamefont
  {Narayan}},\ }\bibfield  {title} {\enquote {\bibinfo {title} {{Small
  Schwarzschild de Sitter black holes, quantum extremal surfaces and
  islands}},}\ }\href {\doibase 10.1007/JHEP10(2022)031} {\bibfield  {journal}
  {\bibinfo  {journal} {JHEP}\ }\textbf {\bibinfo {volume} {10}},\ \bibinfo
  {pages} {031} (\bibinfo {year} {2022})},\ \Eprint
  {http://arxiv.org/abs/2207.10724} {arXiv:2207.10724 [hep-th]} \BibitemShut
  {NoStop}%
\bibitem [{\citenamefont {Chu}\ \emph {et~al.}(2021)\citenamefont {Chu},
  \citenamefont {Deng},\ and\ \citenamefont {Zhou}}]{Chu:2021gdb}%
  \BibitemOpen
  \bibfield  {author} {\bibinfo {author} {\bibfnamefont {Jinwei}\ \bibnamefont
  {Chu}}, \bibinfo {author} {\bibfnamefont {Feiyu}\ \bibnamefont {Deng}}, \
  and\ \bibinfo {author} {\bibfnamefont {Yang}\ \bibnamefont {Zhou}},\
  }\bibfield  {title} {\enquote {\bibinfo {title} {{Page curve from defect
  extremal surface and island in higher dimensions}},}\ }\href {\doibase
  10.1007/JHEP10(2021)149} {\bibfield  {journal} {\bibinfo  {journal} {JHEP}\
  }\textbf {\bibinfo {volume} {10}},\ \bibinfo {pages} {149} (\bibinfo {year}
  {2021})},\ \Eprint {http://arxiv.org/abs/2105.09106} {arXiv:2105.09106
  [hep-th]} \BibitemShut {NoStop}%
\bibitem [{\citenamefont {Ling}\ \emph {et~al.}(2021)\citenamefont {Ling},
  \citenamefont {Liu},\ and\ \citenamefont {Xian}}]{Ling:2020laa}%
  \BibitemOpen
  \bibfield  {author} {\bibinfo {author} {\bibfnamefont {Yi}~\bibnamefont
  {Ling}}, \bibinfo {author} {\bibfnamefont {Yuxuan}\ \bibnamefont {Liu}}, \
  and\ \bibinfo {author} {\bibfnamefont {Zhuo-Yu}\ \bibnamefont {Xian}},\
  }\bibfield  {title} {\enquote {\bibinfo {title} {{Island in Charged Black
  Holes}},}\ }\href {\doibase 10.1007/JHEP03(2021)251} {\bibfield  {journal}
  {\bibinfo  {journal} {JHEP}\ }\textbf {\bibinfo {volume} {03}},\ \bibinfo
  {pages} {251} (\bibinfo {year} {2021})},\ \Eprint
  {http://arxiv.org/abs/2010.00037} {arXiv:2010.00037 [hep-th]} \BibitemShut
  {NoStop}%
\bibitem [{\citenamefont {Wang}\ \emph
  {et~al.}(2021{\natexlab{b}})\citenamefont {Wang}, \citenamefont {Li},\ and\
  \citenamefont {Wang}}]{Wang:2021woy}%
  \BibitemOpen
  \bibfield  {author} {\bibinfo {author} {\bibfnamefont {Xuanhua}\ \bibnamefont
  {Wang}}, \bibinfo {author} {\bibfnamefont {Ran}\ \bibnamefont {Li}}, \ and\
  \bibinfo {author} {\bibfnamefont {Jin}\ \bibnamefont {Wang}},\ }\bibfield
  {title} {\enquote {\bibinfo {title} {{Islands and Page curves of
  Reissner-Nordstr\"om black holes}},}\ }\href {\doibase
  10.1007/JHEP04(2021)103} {\bibfield  {journal} {\bibinfo  {journal} {JHEP}\
  }\textbf {\bibinfo {volume} {04}},\ \bibinfo {pages} {103} (\bibinfo {year}
  {2021}{\natexlab{b}})},\ \Eprint {http://arxiv.org/abs/2101.06867}
  {arXiv:2101.06867 [hep-th]} \BibitemShut {NoStop}%
\bibitem [{\citenamefont {Kim}\ and\ \citenamefont {Nam}(2021)}]{Kim:2021gzd}%
  \BibitemOpen
  \bibfield  {author} {\bibinfo {author} {\bibfnamefont {Wontae}\ \bibnamefont
  {Kim}}\ and\ \bibinfo {author} {\bibfnamefont {Mungon}\ \bibnamefont {Nam}},\
  }\bibfield  {title} {\enquote {\bibinfo {title} {{Entanglement entropy of
  asymptotically flat non-extremal and extremal black holes with an island}},}\
  }\href {\doibase 10.1140/epjc/s10052-021-09680-x} {\bibfield  {journal}
  {\bibinfo  {journal} {Eur. Phys. J. C}\ }\textbf {\bibinfo {volume} {81}},\
  \bibinfo {pages} {869} (\bibinfo {year} {2021})},\ \Eprint
  {http://arxiv.org/abs/2103.16163} {arXiv:2103.16163 [hep-th]} \BibitemShut
  {NoStop}%
\bibitem [{\citenamefont {Ahn}\ \emph {et~al.}(2022)\citenamefont {Ahn},
  \citenamefont {Bak}, \citenamefont {Jeong}, \citenamefont {Kim},\ and\
  \citenamefont {Sun}}]{Ahn:2021chg}%
  \BibitemOpen
  \bibfield  {author} {\bibinfo {author} {\bibfnamefont {Byoungjoon}\
  \bibnamefont {Ahn}}, \bibinfo {author} {\bibfnamefont {Sang-Eon}\
  \bibnamefont {Bak}}, \bibinfo {author} {\bibfnamefont {Hyun-Sik}\
  \bibnamefont {Jeong}}, \bibinfo {author} {\bibfnamefont {Keun-Young}\
  \bibnamefont {Kim}}, \ and\ \bibinfo {author} {\bibfnamefont {Ya-Wen}\
  \bibnamefont {Sun}},\ }\bibfield  {title} {\enquote {\bibinfo {title}
  {{Islands in charged linear dilaton black holes}},}\ }\href {\doibase
  10.1103/PhysRevD.105.046012} {\bibfield  {journal} {\bibinfo  {journal}
  {Phys. Rev. D}\ }\textbf {\bibinfo {volume} {105}},\ \bibinfo {pages}
  {046012} (\bibinfo {year} {2022})},\ \Eprint
  {http://arxiv.org/abs/2107.07444} {arXiv:2107.07444 [hep-th]} \BibitemShut
  {NoStop}%
\bibitem [{\citenamefont {Yu}\ and\ \citenamefont {Ge}(2022)}]{Yu:2021cgi}%
  \BibitemOpen
  \bibfield  {author} {\bibinfo {author} {\bibfnamefont {Ming-Hui}\
  \bibnamefont {Yu}}\ and\ \bibinfo {author} {\bibfnamefont {Xian-Hui}\
  \bibnamefont {Ge}},\ }\bibfield  {title} {\enquote {\bibinfo {title}
  {{Islands and Page curves in charged dilaton black holes}},}\ }\href
  {\doibase 10.1140/epjc/s10052-021-09932-w} {\bibfield  {journal} {\bibinfo
  {journal} {Eur. Phys. J. C}\ }\textbf {\bibinfo {volume} {82}},\ \bibinfo
  {pages} {14} (\bibinfo {year} {2022})},\ \Eprint
  {http://arxiv.org/abs/2107.03031} {arXiv:2107.03031 [hep-th]} \BibitemShut
  {NoStop}%
\bibitem [{\citenamefont {Chen}\ \emph {et~al.}(2021)\citenamefont {Chen},
  \citenamefont {Gorbenko},\ and\ \citenamefont {Maldacena}}]{Chen:2020tes}%
  \BibitemOpen
  \bibfield  {author} {\bibinfo {author} {\bibfnamefont {Yiming}\ \bibnamefont
  {Chen}}, \bibinfo {author} {\bibfnamefont {Victor}\ \bibnamefont {Gorbenko}},
  \ and\ \bibinfo {author} {\bibfnamefont {Juan}\ \bibnamefont {Maldacena}},\
  }\bibfield  {title} {\enquote {\bibinfo {title} {{Bra-ket wormholes in
  gravitationally prepared states}},}\ }\href {\doibase
  10.1007/JHEP02(2021)009} {\bibfield  {journal} {\bibinfo  {journal} {JHEP}\
  }\textbf {\bibinfo {volume} {02}},\ \bibinfo {pages} {009} (\bibinfo {year}
  {2021})},\ \Eprint {http://arxiv.org/abs/2007.16091} {arXiv:2007.16091
  [hep-th]} \BibitemShut {NoStop}%
\bibitem [{\citenamefont {Geng}\ \emph
  {et~al.}(2021{\natexlab{b}})\citenamefont {Geng}, \citenamefont {Nomura},\
  and\ \citenamefont {Sun}}]{Geng:2021wcq}%
  \BibitemOpen
  \bibfield  {author} {\bibinfo {author} {\bibfnamefont {Hao}\ \bibnamefont
  {Geng}}, \bibinfo {author} {\bibfnamefont {Yasunori}\ \bibnamefont {Nomura}},
  \ and\ \bibinfo {author} {\bibfnamefont {Hao-Yu}\ \bibnamefont {Sun}},\
  }\bibfield  {title} {\enquote {\bibinfo {title} {{Information paradox and its
  resolution in de Sitter holography}},}\ }\href {\doibase
  10.1103/PhysRevD.103.126004} {\bibfield  {journal} {\bibinfo  {journal}
  {Phys. Rev. D}\ }\textbf {\bibinfo {volume} {103}},\ \bibinfo {pages}
  {126004} (\bibinfo {year} {2021}{\natexlab{b}})},\ \Eprint
  {http://arxiv.org/abs/2103.07477} {arXiv:2103.07477 [hep-th]} \BibitemShut
  {NoStop}%
\bibitem [{\citenamefont {Bhattacharya}\ \emph {et~al.}(2021)\citenamefont
  {Bhattacharya}, \citenamefont {Bhattacharyya}, \citenamefont {Nandy},\ and\
  \citenamefont {Patra}}]{Bhattacharya:2021jrn}%
  \BibitemOpen
  \bibfield  {author} {\bibinfo {author} {\bibfnamefont {Aranya}\ \bibnamefont
  {Bhattacharya}}, \bibinfo {author} {\bibfnamefont {Arpan}\ \bibnamefont
  {Bhattacharyya}}, \bibinfo {author} {\bibfnamefont {Pratik}\ \bibnamefont
  {Nandy}}, \ and\ \bibinfo {author} {\bibfnamefont {Ayan~K.}\ \bibnamefont
  {Patra}},\ }\bibfield  {title} {\enquote {\bibinfo {title} {{Islands and
  complexity of eternal black hole and radiation subsystems for a doubly
  holographic model}},}\ }\href {\doibase 10.1007/JHEP05(2021)135} {\bibfield
  {journal} {\bibinfo  {journal} {JHEP}\ }\textbf {\bibinfo {volume} {05}},\
  \bibinfo {pages} {135} (\bibinfo {year} {2021})},\ \Eprint
  {http://arxiv.org/abs/2103.15852} {arXiv:2103.15852 [hep-th]} \BibitemShut
  {NoStop}%
\bibitem [{\citenamefont {Yu}\ and\ \citenamefont {Ge}(2024)}]{Yu:2024fks}%
  \BibitemOpen
  \bibfield  {author} {\bibinfo {author} {\bibfnamefont {Ming-Hui}\
  \bibnamefont {Yu}}\ and\ \bibinfo {author} {\bibfnamefont {Xian-Hui}\
  \bibnamefont {Ge}},\ }\bibfield  {title} {\enquote {\bibinfo {title}
  {{Geometric Constraints via Page Curves: Insights from Island Rule and
  Quantum Focusing Conjecture}},}\ }\href@noop {} {\  (\bibinfo {year}
  {2024})},\ \Eprint {http://arxiv.org/abs/2405.03220} {arXiv:2405.03220
  [hep-th]} \BibitemShut {NoStop}%
\bibitem [{\citenamefont {Gomez}(2024)}]{Gomez:2024fij}%
  \BibitemOpen
  \bibfield  {author} {\bibinfo {author} {\bibfnamefont {Cesar}\ \bibnamefont
  {Gomez}},\ }\bibfield  {title} {\enquote {\bibinfo {title} {{The Algebraic
  Page Curve}},}\ }\href@noop {} {\  (\bibinfo {year} {2024})},\ \Eprint
  {http://arxiv.org/abs/2403.09165} {arXiv:2403.09165 [hep-th]} \BibitemShut
  {NoStop}%
\bibitem [{\citenamefont {Jain}\ \emph {et~al.}(2023)\citenamefont {Jain},
  \citenamefont {Pant},\ and\ \citenamefont {Parihar}}]{Jain:2023xta}%
  \BibitemOpen
  \bibfield  {author} {\bibinfo {author} {\bibfnamefont {Parul}\ \bibnamefont
  {Jain}}, \bibinfo {author} {\bibfnamefont {Sanjay}\ \bibnamefont {Pant}}, \
  and\ \bibinfo {author} {\bibfnamefont {Himanshu}\ \bibnamefont {Parihar}},\
  }\bibfield  {title} {\enquote {\bibinfo {title} {{Island in Quark Cloud
  Model}},}\ }\href@noop {} {\  (\bibinfo {year} {2023})},\ \Eprint
  {http://arxiv.org/abs/2311.08186} {arXiv:2311.08186 [hep-th]} \BibitemShut
  {NoStop}%
\bibitem [{\citenamefont {Guo}\ and\ \citenamefont {Miao}(2023)}]{Guo:2023fly}%
  \BibitemOpen
  \bibfield  {author} {\bibinfo {author} {\bibfnamefont {Yu}~\bibnamefont
  {Guo}}\ and\ \bibinfo {author} {\bibfnamefont {Rong-Xin}\ \bibnamefont
  {Miao}},\ }\bibfield  {title} {\enquote {\bibinfo {title} {{Page curves on
  codim-m and charged branes}},}\ }\href {\doibase
  10.1140/epjc/s10052-023-12026-4} {\bibfield  {journal} {\bibinfo  {journal}
  {Eur. Phys. J. C}\ }\textbf {\bibinfo {volume} {83}},\ \bibinfo {pages} {847}
  (\bibinfo {year} {2023})}\BibitemShut {NoStop}%
\bibitem [{\citenamefont {Nikolakopoulou}()}]{2699316}%
  \BibitemOpen
  \bibfield  {author} {\bibinfo {author} {\bibfnamefont {T.}~\bibnamefont
  {Nikolakopoulou}},\ }\emph {\bibinfo {title} {{Wormholes and islands}}},\
  \href@noop {} {\bibinfo {type} {Other thesis}}\BibitemShut {NoStop}%
\bibitem [{\citenamefont {Anand}(2023{\natexlab{a}})}]{Anand:2023ozw}%
  \BibitemOpen
  \bibfield  {author} {\bibinfo {author} {\bibfnamefont {Ankit}\ \bibnamefont
  {Anand}},\ }\bibfield  {title} {\enquote {\bibinfo {title} {{Island in Warped
  AdS Black Holes}},}\ }\href@noop {} {\  (\bibinfo {year}
  {2023}{\natexlab{a}})},\ \Eprint {http://arxiv.org/abs/2308.05432}
  {arXiv:2308.05432 [hep-th]} \BibitemShut {NoStop}%
\bibitem [{\citenamefont {Chang}\ \emph {et~al.}(2023)\citenamefont {Chang},
  \citenamefont {He}, \citenamefont {Liu},\ and\ \citenamefont
  {Zhao}}]{Chang:2023gkt}%
  \BibitemOpen
  \bibfield  {author} {\bibinfo {author} {\bibfnamefont {Jing-Cheng}\
  \bibnamefont {Chang}}, \bibinfo {author} {\bibfnamefont {Song}\ \bibnamefont
  {He}}, \bibinfo {author} {\bibfnamefont {Yu-Xiao}\ \bibnamefont {Liu}}, \
  and\ \bibinfo {author} {\bibfnamefont {Long}\ \bibnamefont {Zhao}},\
  }\bibfield  {title} {\enquote {\bibinfo {title} {{Island formula in Planck
  brane}},}\ }\href {\doibase 10.1007/JHEP11(2023)006} {\bibfield  {journal}
  {\bibinfo  {journal} {JHEP}\ }\textbf {\bibinfo {volume} {11}},\ \bibinfo
  {pages} {006} (\bibinfo {year} {2023})},\ \Eprint
  {http://arxiv.org/abs/2308.03645} {arXiv:2308.03645 [hep-th]} \BibitemShut
  {NoStop}%
\bibitem [{\citenamefont {Li}\ and\ \citenamefont {Miao}(2023)}]{Li:2023fly}%
  \BibitemOpen
  \bibfield  {author} {\bibinfo {author} {\bibfnamefont {Dongqi}\ \bibnamefont
  {Li}}\ and\ \bibinfo {author} {\bibfnamefont {Rong-Xin}\ \bibnamefont
  {Miao}},\ }\bibfield  {title} {\enquote {\bibinfo {title} {{Massless
  entanglement islands in cone holography}},}\ }\href {\doibase
  10.1007/JHEP06(2023)056} {\bibfield  {journal} {\bibinfo  {journal} {JHEP}\
  }\textbf {\bibinfo {volume} {06}},\ \bibinfo {pages} {056} (\bibinfo {year}
  {2023})},\ \Eprint {http://arxiv.org/abs/2303.10958} {arXiv:2303.10958
  [hep-th]} \BibitemShut {NoStop}%
\bibitem [{\citenamefont {Guo}\ \emph {et~al.}(2023)\citenamefont {Guo},
  \citenamefont {Gan},\ and\ \citenamefont {Shu}}]{Guo:2023gfa}%
  \BibitemOpen
  \bibfield  {author} {\bibinfo {author} {\bibfnamefont {Chang-Zhong}\
  \bibnamefont {Guo}}, \bibinfo {author} {\bibfnamefont {Wen-Cong}\
  \bibnamefont {Gan}}, \ and\ \bibinfo {author} {\bibfnamefont {Fu-Wen}\
  \bibnamefont {Shu}},\ }\bibfield  {title} {\enquote {\bibinfo {title} {{Page
  curves and entanglement islands for the step-function Vaidya model of
  evaporating black holes}},}\ }\href {\doibase 10.1007/JHEP05(2023)042}
  {\bibfield  {journal} {\bibinfo  {journal} {JHEP}\ }\textbf {\bibinfo
  {volume} {05}},\ \bibinfo {pages} {042} (\bibinfo {year} {2023})},\ \Eprint
  {http://arxiv.org/abs/2302.02379} {arXiv:2302.02379 [hep-th]} \BibitemShut
  {NoStop}%
\bibitem [{\citenamefont {Miao}(2023)}]{Miao:2023unv}%
  \BibitemOpen
  \bibfield  {author} {\bibinfo {author} {\bibfnamefont {Rong-Xin}\
  \bibnamefont {Miao}},\ }\bibfield  {title} {\enquote {\bibinfo {title}
  {{Entanglement island and Page curve in wedge holography}},}\ }\href
  {\doibase 10.1007/JHEP03(2023)214} {\bibfield  {journal} {\bibinfo  {journal}
  {JHEP}\ }\textbf {\bibinfo {volume} {03}},\ \bibinfo {pages} {214} (\bibinfo
  {year} {2023})},\ \Eprint {http://arxiv.org/abs/2301.06285} {arXiv:2301.06285
  [hep-th]} \BibitemShut {NoStop}%
\bibitem [{\citenamefont {Yadav}\ and\ \citenamefont
  {Joshi}(2023)}]{Yadav:2022jib}%
  \BibitemOpen
  \bibfield  {author} {\bibinfo {author} {\bibfnamefont {Gopal}\ \bibnamefont
  {Yadav}}\ and\ \bibinfo {author} {\bibfnamefont {Nitin}\ \bibnamefont
  {Joshi}},\ }\bibfield  {title} {\enquote {\bibinfo {title} {{Cosmological and
  black hole islands in multi-event horizon spacetimes}},}\ }\href {\doibase
  10.1103/PhysRevD.107.026009} {\bibfield  {journal} {\bibinfo  {journal}
  {Phys. Rev. D}\ }\textbf {\bibinfo {volume} {107}},\ \bibinfo {pages}
  {026009} (\bibinfo {year} {2023})},\ \Eprint
  {http://arxiv.org/abs/2210.00331} {arXiv:2210.00331 [hep-th]} \BibitemShut
  {NoStop}%
\bibitem [{\citenamefont {Hosseini~Mansoori}\ \emph {et~al.}(2022)\citenamefont
  {Hosseini~Mansoori}, \citenamefont {Luongo}, \citenamefont {Mancini},
  \citenamefont {Mirjalali}, \citenamefont {Rafiee},\ and\ \citenamefont
  {Tavanfar}}]{HosseiniMansoori:2022hok}%
  \BibitemOpen
  \bibfield  {author} {\bibinfo {author} {\bibfnamefont {Seyed~Ali}\
  \bibnamefont {Hosseini~Mansoori}}, \bibinfo {author} {\bibfnamefont
  {Orlando}\ \bibnamefont {Luongo}}, \bibinfo {author} {\bibfnamefont
  {Stefano}\ \bibnamefont {Mancini}}, \bibinfo {author} {\bibfnamefont
  {Mirmani}\ \bibnamefont {Mirjalali}}, \bibinfo {author} {\bibfnamefont
  {Morteza}\ \bibnamefont {Rafiee}}, \ and\ \bibinfo {author} {\bibfnamefont
  {Alireza}\ \bibnamefont {Tavanfar}},\ }\bibfield  {title} {\enquote {\bibinfo
  {title} {{Planar black holes in holographic axion gravity: Islands, Page
  times, and scrambling times}},}\ }\href {\doibase
  10.1103/PhysRevD.106.126018} {\bibfield  {journal} {\bibinfo  {journal}
  {Phys. Rev. D}\ }\textbf {\bibinfo {volume} {106}},\ \bibinfo {pages}
  {126018} (\bibinfo {year} {2022})},\ \Eprint
  {http://arxiv.org/abs/2209.00253} {arXiv:2209.00253 [hep-th]} \BibitemShut
  {NoStop}%
\bibitem [{\citenamefont {Hu}\ \emph {et~al.}(2022)\citenamefont {Hu},
  \citenamefont {Li},\ and\ \citenamefont {Miao}}]{Hu:2022zgy}%
  \BibitemOpen
  \bibfield  {author} {\bibinfo {author} {\bibfnamefont {Peng-Ju}\ \bibnamefont
  {Hu}}, \bibinfo {author} {\bibfnamefont {Dongqi}\ \bibnamefont {Li}}, \ and\
  \bibinfo {author} {\bibfnamefont {Rong-Xin}\ \bibnamefont {Miao}},\
  }\bibfield  {title} {\enquote {\bibinfo {title} {{Island on codimension-two
  branes in AdS/dCFT}},}\ }\href {\doibase 10.1007/JHEP11(2022)008} {\bibfield
  {journal} {\bibinfo  {journal} {JHEP}\ }\textbf {\bibinfo {volume} {11}},\
  \bibinfo {pages} {008} (\bibinfo {year} {2022})},\ \Eprint
  {http://arxiv.org/abs/2208.11982} {arXiv:2208.11982 [hep-th]} \BibitemShut
  {NoStop}%
\bibitem [{\citenamefont {Yu}\ and\ \citenamefont {Ge}(2023)}]{Yu:2022xlh}%
  \BibitemOpen
  \bibfield  {author} {\bibinfo {author} {\bibfnamefont {Ming-Hui}\
  \bibnamefont {Yu}}\ and\ \bibinfo {author} {\bibfnamefont {Xian-Hui}\
  \bibnamefont {Ge}},\ }\bibfield  {title} {\enquote {\bibinfo {title}
  {{Entanglement islands in generalized two-dimensional dilaton black
  holes}},}\ }\href {\doibase 10.1103/PhysRevD.107.066020} {\bibfield
  {journal} {\bibinfo  {journal} {Phys. Rev. D}\ }\textbf {\bibinfo {volume}
  {107}},\ \bibinfo {pages} {066020} (\bibinfo {year} {2023})},\ \Eprint
  {http://arxiv.org/abs/2208.01943} {arXiv:2208.01943 [hep-th]} \BibitemShut
  {NoStop}%
\bibitem [{\citenamefont {Djordjevi\'c}\ \emph {et~al.}(2022)\citenamefont
  {Djordjevi\'c}, \citenamefont {Go\v{c}anin}, \citenamefont {Go\v{c}anin},\
  and\ \citenamefont {Radovanovi\'c}}]{Djordjevic:2022qdk}%
  \BibitemOpen
  \bibfield  {author} {\bibinfo {author} {\bibfnamefont {Stefan}\ \bibnamefont
  {Djordjevi\'c}}, \bibinfo {author} {\bibfnamefont {Aleksandra}\ \bibnamefont
  {Go\v{c}anin}}, \bibinfo {author} {\bibfnamefont {Dragoljub}\ \bibnamefont
  {Go\v{c}anin}}, \ and\ \bibinfo {author} {\bibfnamefont {Voja}\ \bibnamefont
  {Radovanovi\'c}},\ }\bibfield  {title} {\enquote {\bibinfo {title} {{Page
  curve for an eternal Schwarzschild black hole in a dimensionally reduced
  model of dilaton gravity}},}\ }\href {\doibase 10.1103/PhysRevD.106.105015}
  {\bibfield  {journal} {\bibinfo  {journal} {Phys. Rev. D}\ }\textbf {\bibinfo
  {volume} {106}},\ \bibinfo {pages} {105015} (\bibinfo {year} {2022})},\
  \Eprint {http://arxiv.org/abs/2207.07409} {arXiv:2207.07409 [hep-th]}
  \BibitemShut {NoStop}%
\bibitem [{\citenamefont {Anand}(2023{\natexlab{b}})}]{Anand:2022mla}%
  \BibitemOpen
  \bibfield  {author} {\bibinfo {author} {\bibfnamefont {Ankit}\ \bibnamefont
  {Anand}},\ }\bibfield  {title} {\enquote {\bibinfo {title} {{Page curve and
  island in EGB gravity}},}\ }\href {\doibase 10.1016/j.nuclphysb.2023.116284}
  {\bibfield  {journal} {\bibinfo  {journal} {Nucl. Phys. B}\ }\textbf
  {\bibinfo {volume} {993}},\ \bibinfo {pages} {116284} (\bibinfo {year}
  {2023}{\natexlab{b}})},\ \Eprint {http://arxiv.org/abs/2205.13785}
  {arXiv:2205.13785 [hep-th]} \BibitemShut {NoStop}%
\bibitem [{\citenamefont {Azarnia}\ and\ \citenamefont
  {Fareghbal}(2022)}]{Azarnia:2022kmp}%
  \BibitemOpen
  \bibfield  {author} {\bibinfo {author} {\bibfnamefont {Sanam}\ \bibnamefont
  {Azarnia}}\ and\ \bibinfo {author} {\bibfnamefont {Reza}\ \bibnamefont
  {Fareghbal}},\ }\bibfield  {title} {\enquote {\bibinfo {title} {{Islands in
  Kerr\textendash{}de Sitter spacetime and their flat limit}},}\ }\href
  {\doibase 10.1103/PhysRevD.106.026012} {\bibfield  {journal} {\bibinfo
  {journal} {Phys. Rev. D}\ }\textbf {\bibinfo {volume} {106}},\ \bibinfo
  {pages} {026012} (\bibinfo {year} {2022})},\ \Eprint
  {http://arxiv.org/abs/2204.08488} {arXiv:2204.08488 [hep-th]} \BibitemShut
  {NoStop}%
\bibitem [{\citenamefont {Seo}(2022)}]{Seo:2022ezk}%
  \BibitemOpen
  \bibfield  {author} {\bibinfo {author} {\bibfnamefont {Min-Seok}\
  \bibnamefont {Seo}},\ }\bibfield  {title} {\enquote {\bibinfo {title}
  {{Information paradox and island in quasi-de Sitter space}},}\ }\href
  {\doibase 10.1140/epjc/s10052-022-11068-4} {\bibfield  {journal} {\bibinfo
  {journal} {Eur. Phys. J. C}\ }\textbf {\bibinfo {volume} {82}},\ \bibinfo
  {pages} {1082} (\bibinfo {year} {2022})},\ \Eprint
  {http://arxiv.org/abs/2204.04585} {arXiv:2204.04585 [hep-th]} \BibitemShut
  {NoStop}%
\bibitem [{\citenamefont {Wu}\ \emph {et~al.}(2024)\citenamefont {Wu},
  \citenamefont {Teng}, \citenamefont {Huang},\ and\ \citenamefont
  {Lu}}]{Wu:2024jgy}%
  \BibitemOpen
  \bibfield  {author} {\bibinfo {author} {\bibfnamefont {Shu-Min}\ \bibnamefont
  {Wu}}, \bibinfo {author} {\bibfnamefont {Xiao-Wei}\ \bibnamefont {Teng}},
  \bibinfo {author} {\bibfnamefont {Xiao-Li}\ \bibnamefont {Huang}}, \ and\
  \bibinfo {author} {\bibfnamefont {Jianbo}\ \bibnamefont {Lu}},\ }\bibfield
  {title} {\enquote {\bibinfo {title} {{Genuine N-partite entanglement in
  Schwarzschild-de Sitter black hole spacetime}},}\ }\href@noop {} {\
  (\bibinfo {year} {2024})},\ \Eprint {http://arxiv.org/abs/2403.11476}
  {arXiv:2403.11476 [gr-qc]} \BibitemShut {NoStop}%
\bibitem [{\citenamefont {Yu}\ \emph {et~al.}(2023)\citenamefont {Yu},
  \citenamefont {Ge},\ and\ \citenamefont {Lu}}]{Yu:2023whl}%
  \BibitemOpen
  \bibfield  {author} {\bibinfo {author} {\bibfnamefont {Ming-Hui}\
  \bibnamefont {Yu}}, \bibinfo {author} {\bibfnamefont {Xian-Hui}\ \bibnamefont
  {Ge}}, \ and\ \bibinfo {author} {\bibfnamefont {Cheng-Yuan}\ \bibnamefont
  {Lu}},\ }\bibfield  {title} {\enquote {\bibinfo {title} {{Page curves for
  accelerating black holes}},}\ }\href {\doibase
  10.1140/epjc/s10052-023-12267-3} {\bibfield  {journal} {\bibinfo  {journal}
  {Eur. Phys. J. C}\ }\textbf {\bibinfo {volume} {83}},\ \bibinfo {pages}
  {1104} (\bibinfo {year} {2023})},\ \Eprint {http://arxiv.org/abs/2306.11407}
  {arXiv:2306.11407 [hep-th]} \BibitemShut {NoStop}%
\bibitem [{\citenamefont {Luongo}\ \emph {et~al.}(2023)\citenamefont {Luongo},
  \citenamefont {Mancini},\ and\ \citenamefont {Pierosara}}]{Luongo:2023jyz}%
  \BibitemOpen
  \bibfield  {author} {\bibinfo {author} {\bibfnamefont {Orlando}\ \bibnamefont
  {Luongo}}, \bibinfo {author} {\bibfnamefont {Stefano}\ \bibnamefont
  {Mancini}}, \ and\ \bibinfo {author} {\bibfnamefont {Paolo}\ \bibnamefont
  {Pierosara}},\ }\bibfield  {title} {\enquote {\bibinfo {title} {{Entanglement
  entropy for spherically symmetric regular black holes}},}\ }\href {\doibase
  10.1103/PhysRevD.108.104059} {\bibfield  {journal} {\bibinfo  {journal}
  {Phys. Rev. D}\ }\textbf {\bibinfo {volume} {108}},\ \bibinfo {pages}
  {104059} (\bibinfo {year} {2023})},\ \Eprint
  {http://arxiv.org/abs/2304.06593} {arXiv:2304.06593 [gr-qc]} \BibitemShut
  {NoStop}%
\bibitem [{\citenamefont {Uhlemann}(2021)}]{Uhlemann_2021}%
  \BibitemOpen
  \bibfield  {author} {\bibinfo {author} {\bibfnamefont {Christoph~F.}\
  \bibnamefont {Uhlemann}},\ }\bibfield  {title} {\enquote {\bibinfo {title}
  {Islands and page curves in 4d from type iib},}\ }\href {\doibase
  10.1007/jhep08(2021)104} {\bibfield  {journal} {\bibinfo  {journal} {Journal
  of High Energy Physics}\ }\textbf {\bibinfo {volume} {2021}} (\bibinfo {year}
  {2021}),\ 10.1007/jhep08(2021)104}\BibitemShut {NoStop}%
\bibitem [{\citenamefont {Karch}\ \emph {et~al.}(2022)\citenamefont {Karch},
  \citenamefont {Sun},\ and\ \citenamefont {Uhlemann}}]{Karch_2022}%
  \BibitemOpen
  \bibfield  {author} {\bibinfo {author} {\bibfnamefont {Andreas}\ \bibnamefont
  {Karch}}, \bibinfo {author} {\bibfnamefont {Hao-Yu}\ \bibnamefont {Sun}}, \
  and\ \bibinfo {author} {\bibfnamefont {Christoph~F.}\ \bibnamefont
  {Uhlemann}},\ }\bibfield  {title} {\enquote {\bibinfo {title} {Double
  holography in string theory},}\ }\href {\doibase 10.1007/jhep10(2022)012}
  {\bibfield  {journal} {\bibinfo  {journal} {Journal of High Energy Physics}\
  }\textbf {\bibinfo {volume} {2022}} (\bibinfo {year} {2022}),\
  10.1007/jhep10(2022)012}\BibitemShut {NoStop}%
\bibitem [{\citenamefont {Saha}\ \emph {et~al.}(2022)\citenamefont {Saha},
  \citenamefont {Gangopadhyay},\ and\ \citenamefont {Saha}}]{Saha:2021ohr}%
  \BibitemOpen
  \bibfield  {author} {\bibinfo {author} {\bibfnamefont {Ashis}\ \bibnamefont
  {Saha}}, \bibinfo {author} {\bibfnamefont {Sunandan}\ \bibnamefont
  {Gangopadhyay}}, \ and\ \bibinfo {author} {\bibfnamefont {Jyoti~Prasad}\
  \bibnamefont {Saha}},\ }\bibfield  {title} {\enquote {\bibinfo {title}
  {{Mutual information, islands in black holes and the Page curve}},}\ }\href
  {\doibase 10.1140/epjc/s10052-022-10426-6} {\bibfield  {journal} {\bibinfo
  {journal} {Eur. Phys. J. C}\ }\textbf {\bibinfo {volume} {82}},\ \bibinfo
  {pages} {476} (\bibinfo {year} {2022})},\ \Eprint
  {http://arxiv.org/abs/2109.02996} {arXiv:2109.02996 [hep-th]} \BibitemShut
  {NoStop}%
\bibitem [{\citenamefont {Roy~Chowdhury}\ \emph {et~al.}(2022)\citenamefont
  {Roy~Chowdhury}, \citenamefont {Saha},\ and\ \citenamefont
  {Gangopadhyay}}]{RoyChowdhury:2022awr}%
  \BibitemOpen
  \bibfield  {author} {\bibinfo {author} {\bibfnamefont {Anirban}\ \bibnamefont
  {Roy~Chowdhury}}, \bibinfo {author} {\bibfnamefont {Ashis}\ \bibnamefont
  {Saha}}, \ and\ \bibinfo {author} {\bibfnamefont {Sunandan}\ \bibnamefont
  {Gangopadhyay}},\ }\bibfield  {title} {\enquote {\bibinfo {title} {{Role of
  mutual information in the Page curve}},}\ }\href {\doibase
  10.1103/PhysRevD.106.086019} {\bibfield  {journal} {\bibinfo  {journal}
  {Phys. Rev. D}\ }\textbf {\bibinfo {volume} {106}},\ \bibinfo {pages}
  {086019} (\bibinfo {year} {2022})},\ \Eprint
  {http://arxiv.org/abs/2207.13029} {arXiv:2207.13029 [hep-th]} \BibitemShut
  {NoStop}%
\bibitem [{\citenamefont {Roy~Chowdhury}\ \emph {et~al.}(2023)\citenamefont
  {Roy~Chowdhury}, \citenamefont {Saha},\ and\ \citenamefont
  {Gangopadhyay}}]{RoyChowdhury:2023eol}%
  \BibitemOpen
  \bibfield  {author} {\bibinfo {author} {\bibfnamefont {Anirban}\ \bibnamefont
  {Roy~Chowdhury}}, \bibinfo {author} {\bibfnamefont {Ashis}\ \bibnamefont
  {Saha}}, \ and\ \bibinfo {author} {\bibfnamefont {Sunandan}\ \bibnamefont
  {Gangopadhyay}},\ }\bibfield  {title} {\enquote {\bibinfo {title} {{Mutual
  information of subsystems and the Page curve for the
  Schwarzschild\textendash{}de Sitter black hole}},}\ }\href {\doibase
  10.1103/PhysRevD.108.026003} {\bibfield  {journal} {\bibinfo  {journal}
  {Phys. Rev. D}\ }\textbf {\bibinfo {volume} {108}},\ \bibinfo {pages}
  {026003} (\bibinfo {year} {2023})},\ \Eprint
  {http://arxiv.org/abs/2303.14062} {arXiv:2303.14062 [hep-th]} \BibitemShut
  {NoStop}%
\bibitem [{\citenamefont {Reuter}(1998)}]{Reuter:1996cp}%
  \BibitemOpen
  \bibfield  {author} {\bibinfo {author} {\bibfnamefont {M.}~\bibnamefont
  {Reuter}},\ }\bibfield  {title} {\enquote {\bibinfo {title} {{Nonperturbative
  evolution equation for quantum gravity}},}\ }\href {\doibase
  10.1103/PhysRevD.57.971} {\bibfield  {journal} {\bibinfo  {journal} {Phys.
  Rev. D}\ }\textbf {\bibinfo {volume} {57}},\ \bibinfo {pages} {971--985}
  (\bibinfo {year} {1998})},\ \Eprint {http://arxiv.org/abs/hep-th/9605030}
  {arXiv:hep-th/9605030} \BibitemShut {NoStop}%
\bibitem [{\citenamefont {Percacci}(2017)}]{doi:10.1142/10369}%
  \BibitemOpen
  \bibfield  {author} {\bibinfo {author} {\bibfnamefont {Roberto}\ \bibnamefont
  {Percacci}},\ }\href {\doibase 10.1142/10369} {\emph {\bibinfo {title} {An
  Introduction to Covariant Quantum Gravity and Asymptotic Safety}}}\ (\bibinfo
   {publisher} {World Scientific},\ \bibinfo {year} {2017})\ \Eprint
  {http://arxiv.org/abs/https://www.worldscientific.com/doi/pdf/10.1142/10369}
  {https://www.worldscientific.com/doi/pdf/10.1142/10369} \BibitemShut
  {NoStop}%
\bibitem [{\citenamefont {Reuter}\ and\ \citenamefont
  {Saueressig}(2019)}]{reuter2019quantum}%
  \BibitemOpen
  \bibfield  {author} {\bibinfo {author} {\bibfnamefont {M.}~\bibnamefont
  {Reuter}}\ and\ \bibinfo {author} {\bibfnamefont {F.}~\bibnamefont
  {Saueressig}},\ }\href {https://books.google.co.in/books?id=WKh7DwAAQBAJ}
  {\emph {\bibinfo {title} {Quantum Gravity and the Functional Renormalization
  Group: The Road towards Asymptotic Safety}}},\ Cambridge Monographs on
  Mathematical Physics\ (\bibinfo  {publisher} {Cambridge University Press},\
  \bibinfo {year} {2019})\BibitemShut {NoStop}%
\bibitem [{\citenamefont {Weinberg}(1980)}]{Weinberg:1980gg}%
  \BibitemOpen
  \bibfield  {author} {\bibinfo {author} {\bibfnamefont {Steven}\ \bibnamefont
  {Weinberg}},\ }\enquote {\bibinfo {title} {{ULTRAVIOLET DIVERGENCES IN
  QUANTUM THEORIES OF GRAVITATION}},}\ in\ \href@noop {} {\emph {\bibinfo
  {booktitle} {{General Relativity}: {An Einstein Centenary Survey}}}}\
  (\bibinfo {year} {1980})\ pp.\ \bibinfo {pages} {790--831}\BibitemShut
  {NoStop}%
\bibitem [{\citenamefont {Reuter}\ and\ \citenamefont
  {Saueressig}(2002)}]{Reuter:2001ag}%
  \BibitemOpen
  \bibfield  {author} {\bibinfo {author} {\bibfnamefont {M.}~\bibnamefont
  {Reuter}}\ and\ \bibinfo {author} {\bibfnamefont {Frank}\ \bibnamefont
  {Saueressig}},\ }\bibfield  {title} {\enquote {\bibinfo {title}
  {{Renormalization group flow of quantum gravity in the Einstein-Hilbert
  truncation}},}\ }\href {\doibase 10.1103/PhysRevD.65.065016} {\bibfield
  {journal} {\bibinfo  {journal} {Phys. Rev. D}\ }\textbf {\bibinfo {volume}
  {65}},\ \bibinfo {pages} {065016} (\bibinfo {year} {2002})},\ \Eprint
  {http://arxiv.org/abs/hep-th/0110054} {arXiv:hep-th/0110054} \BibitemShut
  {NoStop}%
\bibitem [{\citenamefont {Wetterich}(1993)}]{Wetterich:1992yh}%
  \BibitemOpen
  \bibfield  {author} {\bibinfo {author} {\bibfnamefont {Christof}\
  \bibnamefont {Wetterich}},\ }\bibfield  {title} {\enquote {\bibinfo {title}
  {{Exact evolution equation for the effective potential}},}\ }\href {\doibase
  10.1016/0370-2693(93)90726-X} {\bibfield  {journal} {\bibinfo  {journal}
  {Phys. Lett. B}\ }\textbf {\bibinfo {volume} {301}},\ \bibinfo {pages}
  {90--94} (\bibinfo {year} {1993})},\ \Eprint
  {http://arxiv.org/abs/1710.05815} {arXiv:1710.05815 [hep-th]} \BibitemShut
  {NoStop}%
\bibitem [{\citenamefont {Reuter}\ and\ \citenamefont
  {Wetterich}(1994)}]{Reuter:1993kw}%
  \BibitemOpen
  \bibfield  {author} {\bibinfo {author} {\bibfnamefont {M.}~\bibnamefont
  {Reuter}}\ and\ \bibinfo {author} {\bibfnamefont {C.}~\bibnamefont
  {Wetterich}},\ }\bibfield  {title} {\enquote {\bibinfo {title} {{Effective
  average action for gauge theories and exact evolution equations}},}\ }\href
  {\doibase 10.1016/0550-3213(94)90543-6} {\bibfield  {journal} {\bibinfo
  {journal} {Nucl. Phys. B}\ }\textbf {\bibinfo {volume} {417}},\ \bibinfo
  {pages} {181--214} (\bibinfo {year} {1994})}\BibitemShut {NoStop}%
\bibitem [{\citenamefont {Lauscher}\ and\ \citenamefont
  {Reuter}(2002)}]{Lauscher:2002sq}%
  \BibitemOpen
  \bibfield  {author} {\bibinfo {author} {\bibfnamefont {O.}~\bibnamefont
  {Lauscher}}\ and\ \bibinfo {author} {\bibfnamefont {M.}~\bibnamefont
  {Reuter}},\ }\bibfield  {title} {\enquote {\bibinfo {title} {{Flow equation
  of quantum Einstein gravity in a higher derivative truncation}},}\ }\href
  {\doibase 10.1103/PhysRevD.66.025026} {\bibfield  {journal} {\bibinfo
  {journal} {Phys. Rev. D}\ }\textbf {\bibinfo {volume} {66}},\ \bibinfo
  {pages} {025026} (\bibinfo {year} {2002})},\ \Eprint
  {http://arxiv.org/abs/hep-th/0205062} {arXiv:hep-th/0205062} \BibitemShut
  {NoStop}%
\bibitem [{\citenamefont {Niedermaier}(2009)}]{PhysRevLett.103.101303}%
  \BibitemOpen
  \bibfield  {author} {\bibinfo {author} {\bibfnamefont {Max~R.}\ \bibnamefont
  {Niedermaier}},\ }\bibfield  {title} {\enquote {\bibinfo {title}
  {Gravitational fixed points from perturbation theory},}\ }\href {\doibase
  10.1103/PhysRevLett.103.101303} {\bibfield  {journal} {\bibinfo  {journal}
  {Phys. Rev. Lett.}\ }\textbf {\bibinfo {volume} {103}},\ \bibinfo {pages}
  {101303} (\bibinfo {year} {2009})}\BibitemShut {NoStop}%
\bibitem [{\citenamefont {Mandal}\ \emph {et~al.}(2020)\citenamefont {Mandal},
  \citenamefont {Gangopadhyay},\ and\ \citenamefont {Lahiri}}]{Mandal:2019xlg}%
  \BibitemOpen
  \bibfield  {author} {\bibinfo {author} {\bibfnamefont {Rituparna}\
  \bibnamefont {Mandal}}, \bibinfo {author} {\bibfnamefont {Sunandan}\
  \bibnamefont {Gangopadhyay}}, \ and\ \bibinfo {author} {\bibfnamefont
  {Amitabha}\ \bibnamefont {Lahiri}},\ }\bibfield  {title} {\enquote {\bibinfo
  {title} {{Cosmology of Bianchi type-I metric using renormalization group
  approach for quantum gravity}},}\ }\href {\doibase 10.1088/1361-6382/ab7287}
  {\bibfield  {journal} {\bibinfo  {journal} {Class. Quant. Grav.}\ }\textbf
  {\bibinfo {volume} {37}},\ \bibinfo {pages} {065012} (\bibinfo {year}
  {2020})},\ \Eprint {http://arxiv.org/abs/1906.08674} {arXiv:1906.08674
  [gr-qc]} \BibitemShut {NoStop}%
\bibitem [{\citenamefont {Mandal}\ \emph {et~al.}(2022)\citenamefont {Mandal},
  \citenamefont {Gangopadhyay},\ and\ \citenamefont {Lahiri}}]{Mandal:2020umo}%
  \BibitemOpen
  \bibfield  {author} {\bibinfo {author} {\bibfnamefont {Rituparna}\
  \bibnamefont {Mandal}}, \bibinfo {author} {\bibfnamefont {Sunandan}\
  \bibnamefont {Gangopadhyay}}, \ and\ \bibinfo {author} {\bibfnamefont
  {Amitabha}\ \bibnamefont {Lahiri}},\ }\bibfield  {title} {\enquote {\bibinfo
  {title} {{Cosmology with modified continuity equation in asymptotically safe
  gravity}},}\ }\href@noop {} {\bibfield  {journal} {\bibinfo  {journal} {Eur.
  Phys. J. Plus}\ }\textbf {\bibinfo {volume} {137}},\ \bibinfo {pages} {10}
  (\bibinfo {year} {2022})},\ \Eprint {http://arxiv.org/abs/2010.09716}
  {arXiv:2010.09716 [gr-qc]} \BibitemShut {NoStop}%
\bibitem [{\citenamefont {Niedermaier}(2007)}]{Niedermaier:2006ns}%
  \BibitemOpen
  \bibfield  {author} {\bibinfo {author} {\bibfnamefont {M.}~\bibnamefont
  {Niedermaier}},\ }\bibfield  {title} {\enquote {\bibinfo {title} {{The
  Asymptotic safety scenario in quantum gravity: An Introduction}},}\ }\href
  {\doibase 10.1088/0264-9381/24/18/R01} {\bibfield  {journal} {\bibinfo
  {journal} {Class. Quant. Grav.}\ }\textbf {\bibinfo {volume} {24}},\ \bibinfo
  {pages} {R171--230} (\bibinfo {year} {2007})},\ \Eprint
  {http://arxiv.org/abs/gr-qc/0610018} {arXiv:gr-qc/0610018} \BibitemShut
  {NoStop}%
\bibitem [{\citenamefont {Niedermaier}(2003)}]{Niedermaier:2003fz}%
  \BibitemOpen
  \bibfield  {author} {\bibinfo {author} {\bibfnamefont {Max}\ \bibnamefont
  {Niedermaier}},\ }\bibfield  {title} {\enquote {\bibinfo {title}
  {{Dimensionally reduced gravity theories are asymptotically safe}},}\ }\href
  {\doibase 10.1016/j.nuclphysb.2003.09.015} {\bibfield  {journal} {\bibinfo
  {journal} {Nucl. Phys. B}\ }\textbf {\bibinfo {volume} {673}},\ \bibinfo
  {pages} {131--169} (\bibinfo {year} {2003})},\ \Eprint
  {http://arxiv.org/abs/hep-th/0304117} {arXiv:hep-th/0304117} \BibitemShut
  {NoStop}%
\bibitem [{\citenamefont {Nink}\ and\ \citenamefont
  {Reuter}(2013)}]{Nink:2012vd}%
  \BibitemOpen
  \bibfield  {author} {\bibinfo {author} {\bibfnamefont {Andreas}\ \bibnamefont
  {Nink}}\ and\ \bibinfo {author} {\bibfnamefont {Martin}\ \bibnamefont
  {Reuter}},\ }\bibfield  {title} {\enquote {\bibinfo {title} {{On the physical
  mechanism underlying Asymptotic Safety}},}\ }\href {\doibase
  10.1007/JHEP01(2013)062} {\bibfield  {journal} {\bibinfo  {journal} {JHEP}\
  }\textbf {\bibinfo {volume} {01}},\ \bibinfo {pages} {062} (\bibinfo {year}
  {2013})},\ \Eprint {http://arxiv.org/abs/1208.0031} {arXiv:1208.0031
  [hep-th]} \BibitemShut {NoStop}%
\bibitem [{\citenamefont {Litim}\ and\ \citenamefont
  {Sannino}(2014)}]{Litim:2014uca}%
  \BibitemOpen
  \bibfield  {author} {\bibinfo {author} {\bibfnamefont {Daniel~F.}\
  \bibnamefont {Litim}}\ and\ \bibinfo {author} {\bibfnamefont {Francesco}\
  \bibnamefont {Sannino}},\ }\bibfield  {title} {\enquote {\bibinfo {title}
  {{Asymptotic safety guaranteed}},}\ }\href {\doibase 10.1007/JHEP12(2014)178}
  {\bibfield  {journal} {\bibinfo  {journal} {JHEP}\ }\textbf {\bibinfo
  {volume} {12}},\ \bibinfo {pages} {178} (\bibinfo {year} {2014})},\ \Eprint
  {http://arxiv.org/abs/1406.2337} {arXiv:1406.2337 [hep-th]} \BibitemShut
  {NoStop}%
\bibitem [{\citenamefont {Bonanno}\ and\ \citenamefont
  {Reuter}(2000)}]{Bonanno:2000ep}%
  \BibitemOpen
  \bibfield  {author} {\bibinfo {author} {\bibfnamefont {Alfio}\ \bibnamefont
  {Bonanno}}\ and\ \bibinfo {author} {\bibfnamefont {Martin}\ \bibnamefont
  {Reuter}},\ }\bibfield  {title} {\enquote {\bibinfo {title} {{Renormalization
  group improved black hole space-times}},}\ }\href {\doibase
  10.1103/PhysRevD.62.043008} {\bibfield  {journal} {\bibinfo  {journal} {Phys.
  Rev. D}\ }\textbf {\bibinfo {volume} {62}},\ \bibinfo {pages} {043008}
  (\bibinfo {year} {2000})},\ \Eprint {http://arxiv.org/abs/hep-th/0002196}
  {arXiv:hep-th/0002196} \BibitemShut {NoStop}%
\bibitem [{\citenamefont {Mandal}\ and\ \citenamefont
  {Gangopadhyay}(2022)}]{Mandal:2022quv}%
  \BibitemOpen
  \bibfield  {author} {\bibinfo {author} {\bibfnamefont {Rituparna}\
  \bibnamefont {Mandal}}\ and\ \bibinfo {author} {\bibfnamefont {Sunandan}\
  \bibnamefont {Gangopadhyay}},\ }\bibfield  {title} {\enquote {\bibinfo
  {title} {{Black hole thermodynamics in asymptotically safe gravity}},}\
  }\href {\doibase 10.1007/s10714-022-03045-9} {\bibfield  {journal} {\bibinfo
  {journal} {Gen. Rel. Grav.}\ }\textbf {\bibinfo {volume} {54}},\ \bibinfo
  {pages} {159} (\bibinfo {year} {2022})},\ \Eprint
  {http://arxiv.org/abs/2204.11616} {arXiv:2204.11616 [gr-qc]} \BibitemShut
  {NoStop}%
\bibitem [{\citenamefont {Koch}\ and\ \citenamefont
  {Saueressig}(2014)}]{Koch:2014cqa}%
  \BibitemOpen
  \bibfield  {author} {\bibinfo {author} {\bibfnamefont {Benjamin}\
  \bibnamefont {Koch}}\ and\ \bibinfo {author} {\bibfnamefont {Frank}\
  \bibnamefont {Saueressig}},\ }\bibfield  {title} {\enquote {\bibinfo {title}
  {{Black holes within Asymptotic Safety}},}\ }\href {\doibase
  10.1142/S0217751X14300117} {\bibfield  {journal} {\bibinfo  {journal} {Int.
  J. Mod. Phys. A}\ }\textbf {\bibinfo {volume} {29}},\ \bibinfo {pages}
  {1430011} (\bibinfo {year} {2014})},\ \Eprint
  {http://arxiv.org/abs/1401.4452} {arXiv:1401.4452 [hep-th]} \BibitemShut
  {NoStop}%
\bibitem [{\citenamefont {Alishahiha}\ \emph
  {et~al.}(2021{\natexlab{b}})\citenamefont {Alishahiha}, \citenamefont
  {Astaneh},\ and\ \citenamefont {Naseh}}]{Alishahiha_2021}%
  \BibitemOpen
  \bibfield  {author} {\bibinfo {author} {\bibfnamefont {Mohsen}\ \bibnamefont
  {Alishahiha}}, \bibinfo {author} {\bibfnamefont {Amin~Faraji}\ \bibnamefont
  {Astaneh}}, \ and\ \bibinfo {author} {\bibfnamefont {Ali}\ \bibnamefont
  {Naseh}},\ }\bibfield  {title} {\enquote {\bibinfo {title} {Island in the
  presence of higher derivative terms},}\ }\href {\doibase
  10.1007/jhep02(2021)035} {\bibfield  {journal} {\bibinfo  {journal} {Journal
  of High Energy Physics}\ }\textbf {\bibinfo {volume} {2021}} (\bibinfo {year}
  {2021}{\natexlab{b}}),\ 10.1007/jhep02(2021)035}\BibitemShut {NoStop}%
\bibitem [{\citenamefont {Tong}\ \emph {et~al.}(2024)\citenamefont {Tong},
  \citenamefont {Du},\ and\ \citenamefont {Sun}}]{Tong:2023nvi}%
  \BibitemOpen
  \bibfield  {author} {\bibinfo {author} {\bibfnamefont {Chen-Wei}\
  \bibnamefont {Tong}}, \bibinfo {author} {\bibfnamefont {Dong-Hui}\
  \bibnamefont {Du}}, \ and\ \bibinfo {author} {\bibfnamefont {Jia-Rui}\
  \bibnamefont {Sun}},\ }\bibfield  {title} {\enquote {\bibinfo {title}
  {{Island of Reissner-Nordstr{\"o}m anti{\textendash}de Sitter black holes in
  the large D limit}},}\ }\href {\doibase 10.1103/PhysRevD.109.104053}
  {\bibfield  {journal} {\bibinfo  {journal} {Phys. Rev. D}\ }\textbf {\bibinfo
  {volume} {109}},\ \bibinfo {pages} {104053} (\bibinfo {year} {2024})},\
  \Eprint {http://arxiv.org/abs/2306.06682} {arXiv:2306.06682 [hep-th]}
  \BibitemShut {NoStop}%
\bibitem [{\citenamefont {Yadav}\ and\ \citenamefont
  {Misra}(2023)}]{Yadav:2022mnv}%
  \BibitemOpen
  \bibfield  {author} {\bibinfo {author} {\bibfnamefont {Gopal}\ \bibnamefont
  {Yadav}}\ and\ \bibinfo {author} {\bibfnamefont {Aalok}\ \bibnamefont
  {Misra}},\ }\bibfield  {title} {\enquote {\bibinfo {title} {{Entanglement
  entropy and Page curve from the M-theory dual of thermal QCD above Tc at
  intermediate coupling}},}\ }\href {\doibase 10.1103/PhysRevD.107.106015}
  {\bibfield  {journal} {\bibinfo  {journal} {Phys. Rev. D}\ }\textbf {\bibinfo
  {volume} {107}},\ \bibinfo {pages} {106015} (\bibinfo {year} {2023})},\
  \Eprint {http://arxiv.org/abs/2207.04048} {arXiv:2207.04048 [hep-th]}
  \BibitemShut {NoStop}%
\bibitem [{\citenamefont {Liu}\ \emph {et~al.}(2025)\citenamefont {Liu},
  \citenamefont {Xu},\ and\ \citenamefont {Zhang}}]{Liu:2025flo}%
  \BibitemOpen
  \bibfield  {author} {\bibinfo {author} {\bibfnamefont {Yipeng}\ \bibnamefont
  {Liu}}, \bibinfo {author} {\bibfnamefont {Wei}\ \bibnamefont {Xu}}, \ and\
  \bibinfo {author} {\bibfnamefont {Baocheng}\ \bibnamefont {Zhang}},\
  }\bibfield  {title} {\enquote {\bibinfo {title} {{Island rules for the
  noncommutative black hole}},}\ }\href {\doibase
  10.1016/j.physletb.2025.139546} {\bibfield  {journal} {\bibinfo  {journal}
  {Phys. Lett. B}\ }\textbf {\bibinfo {volume} {866}},\ \bibinfo {pages}
  {139546} (\bibinfo {year} {2025})},\ \Eprint
  {http://arxiv.org/abs/2505.09157} {arXiv:2505.09157 [hep-th]} \BibitemShut
  {NoStop}%
\bibitem [{\citenamefont {Dong}\ and\ \citenamefont
  {Tian}(2025)}]{Dong:2025duz}%
  \BibitemOpen
  \bibfield  {author} {\bibinfo {author} {\bibfnamefont {Chen-Yang}\
  \bibnamefont {Dong}}\ and\ \bibinfo {author} {\bibfnamefont {Li-Jun}\
  \bibnamefont {Tian}},\ }\bibfield  {title} {\enquote {\bibinfo {title}
  {{Hawking radiation and Page curve of regular black holes}},}\ }\href
  {\doibase 10.1088/1674-1056/ad989b} {\bibfield  {journal} {\bibinfo
  {journal} {Chin. Phys. B}\ }\textbf {\bibinfo {volume} {34}},\ \bibinfo
  {pages} {020401} (\bibinfo {year} {2025})}\BibitemShut {NoStop}%
\bibitem [{\citenamefont {Lu}\ and\ \citenamefont {Lin}(2022)}]{Lu:2021gmv}%
  \BibitemOpen
  \bibfield  {author} {\bibinfo {author} {\bibfnamefont {Yizhou}\ \bibnamefont
  {Lu}}\ and\ \bibinfo {author} {\bibfnamefont {Jiong}\ \bibnamefont {Lin}},\
  }\bibfield  {title} {\enquote {\bibinfo {title} {{Islands in
  Kaluza{\textendash}Klein black holes}},}\ }\href {\doibase
  10.1140/epjc/s10052-022-10074-w} {\bibfield  {journal} {\bibinfo  {journal}
  {Eur. Phys. J. C}\ }\textbf {\bibinfo {volume} {82}},\ \bibinfo {pages} {132}
  (\bibinfo {year} {2022})},\ \Eprint {http://arxiv.org/abs/2106.07845}
  {arXiv:2106.07845 [hep-th]} \BibitemShut {NoStop}%
\bibitem [{\citenamefont {Calabrese}\ \emph {et~al.}(2009)\citenamefont
  {Calabrese}, \citenamefont {Cardy},\ and\ \citenamefont
  {Tonni}}]{Calabrese:2009ez}%
  \BibitemOpen
  \bibfield  {author} {\bibinfo {author} {\bibfnamefont {P.}~\bibnamefont
  {Calabrese}}, \bibinfo {author} {\bibfnamefont {J.}~\bibnamefont {Cardy}}, \
  and\ \bibinfo {author} {\bibfnamefont {E.}~\bibnamefont {Tonni}},\ }\bibfield
   {title} {\enquote {\bibinfo {title} {{Entanglement entropy of two disjoint
  intervals in conformal field theory}},}\ }\href {\doibase
  10.1088/1742-5468/2009/11/P11001} {\bibfield  {journal} {\bibinfo  {journal}
  {J. Stat. Mech.}\ }\textbf {\bibinfo {volume} {0911}},\ \bibinfo {pages}
  {P11001} (\bibinfo {year} {2009})},\ \Eprint {http://arxiv.org/abs/0905.2069}
  {arXiv:0905.2069 [hep-th]} \BibitemShut {NoStop}%
\bibitem [{\citenamefont {Sekino}\ and\ \citenamefont
  {Susskind}(2008)}]{Sekino:2008he}%
  \BibitemOpen
  \bibfield  {author} {\bibinfo {author} {\bibfnamefont {Y.}~\bibnamefont
  {Sekino}}\ and\ \bibinfo {author} {\bibfnamefont {L.}~\bibnamefont
  {Susskind}},\ }\bibfield  {title} {\enquote {\bibinfo {title} {{Fast
  Scramblers}},}\ }\href {\doibase 10.1088/JHEP10(2008)065} {\bibfield
  {journal} {\bibinfo  {journal} {JHEP}\ }\textbf {\bibinfo {volume} {10}},\
  \bibinfo {pages} {065} (\bibinfo {year} {2008})},\ \Eprint
  {http://arxiv.org/abs/0808.2096} {arXiv:0808.2096 [hep-th]} \BibitemShut
  {NoStop}%
\bibitem [{\citenamefont {Hayden}\ and\ \citenamefont
  {Preskill}(2007)}]{Hayden:2007cs}%
  \BibitemOpen
  \bibfield  {author} {\bibinfo {author} {\bibfnamefont {P.}~\bibnamefont
  {Hayden}}\ and\ \bibinfo {author} {\bibfnamefont {J.}~\bibnamefont
  {Preskill}},\ }\bibfield  {title} {\enquote {\bibinfo {title} {{Black holes
  as mirrors: Quantum information in random subsystems}},}\ }\href {\doibase
  10.1088/1126-6708/2007/09/120} {\bibfield  {journal} {\bibinfo  {journal}
  {JHEP}\ }\textbf {\bibinfo {volume} {09}},\ \bibinfo {pages} {120} (\bibinfo
  {year} {2007})},\ \Eprint {http://arxiv.org/abs/0708.4025} {arXiv:0708.4025
  [hep-th]} \BibitemShut {NoStop}%
\bibitem [{\citenamefont {Takayanagi}\ and\ \citenamefont
  {Umemoto}(2018)}]{Takayanagi:2017knl}%
  \BibitemOpen
  \bibfield  {author} {\bibinfo {author} {\bibfnamefont {Tadashi}\ \bibnamefont
  {Takayanagi}}\ and\ \bibinfo {author} {\bibfnamefont {Koji}\ \bibnamefont
  {Umemoto}},\ }\bibfield  {title} {\enquote {\bibinfo {title} {{Entanglement
  of purification through holographic duality}},}\ }\href {\doibase
  10.1038/s41567-018-0075-2} {\bibfield  {journal} {\bibinfo  {journal} {Nature
  Phys.}\ }\textbf {\bibinfo {volume} {14}},\ \bibinfo {pages} {573--577}
  (\bibinfo {year} {2018})},\ \Eprint {http://arxiv.org/abs/1708.09393}
  {arXiv:1708.09393 [hep-th]} \BibitemShut {NoStop}%
\bibitem [{\citenamefont {Calabrese}\ and\ \citenamefont
  {Cardy}(2009)}]{Calabrese:2009qy}%
  \BibitemOpen
  \bibfield  {author} {\bibinfo {author} {\bibfnamefont {Pasquale}\
  \bibnamefont {Calabrese}}\ and\ \bibinfo {author} {\bibfnamefont {John}\
  \bibnamefont {Cardy}},\ }\bibfield  {title} {\enquote {\bibinfo {title}
  {{Entanglement entropy and conformal field theory}},}\ }\href {\doibase
  10.1088/1751-8113/42/50/504005} {\bibfield  {journal} {\bibinfo  {journal}
  {J. Phys. A}\ }\textbf {\bibinfo {volume} {42}},\ \bibinfo {pages} {504005}
  (\bibinfo {year} {2009})},\ \Eprint {http://arxiv.org/abs/0905.4013}
  {arXiv:0905.4013 [cond-mat.stat-mech]} \BibitemShut {NoStop}%
\bibitem [{\citenamefont {Hartman}\ and\ \citenamefont
  {Maldacena}(2013)}]{Hartman:2013qma}%
  \BibitemOpen
  \bibfield  {author} {\bibinfo {author} {\bibfnamefont {Thomas}\ \bibnamefont
  {Hartman}}\ and\ \bibinfo {author} {\bibfnamefont {Juan}\ \bibnamefont
  {Maldacena}},\ }\bibfield  {title} {\enquote {\bibinfo {title} {{Time
  Evolution of Entanglement Entropy from Black Hole Interiors}},}\ }\href
  {\doibase 10.1007/JHEP05(2013)014} {\bibfield  {journal} {\bibinfo  {journal}
  {JHEP}\ }\textbf {\bibinfo {volume} {05}},\ \bibinfo {pages} {014} (\bibinfo
  {year} {2013})},\ \Eprint {http://arxiv.org/abs/1303.1080} {arXiv:1303.1080
  [hep-th]} \BibitemShut {NoStop}%
\bibitem [{\citenamefont {Grimaldi}\ \emph {et~al.}(2022)\citenamefont
  {Grimaldi}, \citenamefont {Hernandez},\ and\ \citenamefont
  {Myers}}]{Grimaldi:2022suv}%
  \BibitemOpen
  \bibfield  {author} {\bibinfo {author} {\bibfnamefont {Guglielmo}\
  \bibnamefont {Grimaldi}}, \bibinfo {author} {\bibfnamefont {Juan}\
  \bibnamefont {Hernandez}}, \ and\ \bibinfo {author} {\bibfnamefont
  {Robert~C.}\ \bibnamefont {Myers}},\ }\bibfield  {title} {\enquote {\bibinfo
  {title} {{Quantum extremal islands made easy. Part IV. Massive black holes on
  the brane}},}\ }\href {\doibase 10.1007/JHEP03(2022)136} {\bibfield
  {journal} {\bibinfo  {journal} {JHEP}\ }\textbf {\bibinfo {volume} {03}},\
  \bibinfo {pages} {136} (\bibinfo {year} {2022})},\ \Eprint
  {http://arxiv.org/abs/2202.00679} {arXiv:2202.00679 [hep-th]} \BibitemShut
  {NoStop}%
\end{thebibliography}%
\end{document}